\documentclass[a4paper,12pt]{article}
\pdfoutput=0

\usepackage{amsmath}
\usepackage{amsfonts}
\usepackage{amssymb}
\usepackage{graphicx}
\def\be{\begin{equation}}
\def\ee{\end{equation}}
\def\beq{\begin{equation}}
\def\eeq{\end{equation}}

\newcommand{\rmd}{\mathrm{d}}

\newcommand{\Mpl}{M_{\textrm{Pl}}}

\newcommand{\z}{\zeta}
\newcommand{\zb}{\overline{\zeta}}

\renewcommand\[{\left[}
\renewcommand\]{\right]}

\begin{document}

\begin{center}
\Large{\textbf{Conformal consistency relations for single-field inflation}} \\[0.5cm]
 
\large{Paolo Creminelli$^{\rm a}$, Jorge Nore\~na$^{\rm b}$ and Marko Simonovi\'c$^{\rm c}$ }
\\[0.5cm]

\small{
\textit{$^{\rm a}$ Abdus Salam International Centre for Theoretical Physics\\ Strada Costiera 11, 34151, Trieste, Italy}}

\vspace{.2cm}

\small{
\textit{$^{\rm b}$ Institut de Ci\`encies del Cosmos (ICC), \\
Universitat de Barcelona, Mart\'i i Franqu\`es 1, E08028- Spain}}

\vspace{.2cm}

\small{
\textit{$^{\rm c}$ SISSA, via Bonomea 265, 34136, Trieste, Italy}}

\vspace{.2cm}

\end{center}

\vspace{.8cm}

\hrule \vspace{0.3cm}
\noindent \small{\textbf{Abstract}\\ 
We generalize the single-field consistency relations to capture not only the leading term in the squeezed limit---going as $1/q^3$, where $q$ is the small wavevector---but also the subleading one, going as $1/q^2$. This term, for an $(n+1)$-point function, is fixed in terms of the variation of the $n$-point function under a special conformal transformation; this parallels the fact that the $1/q^3$ term is related with the scale dependence of the $n$-point function. 
For the squeezed limit of the 3-point function, this conformal consistency relation implies that there are no terms going as $1/q^2$. We verify that the squeezed limit of the 4-point function is related to the conformal variation of the 3-point function both in the case of canonical slow-roll inflation and in models with  reduced speed of sound. In the second case the conformal consistency conditions capture, at the level of observables, the relation among operators induced by the non-linear realization of Lorentz invariance in the Lagrangian. These results mean that, in any single-field model, primordial correlation functions of $\zeta$ are endowed with an $SO(4,1)$ symmetry, with dilations and special conformal transformations non-linearly realized by $\zeta$.
We also verify the conformal consistency relations for any $n$-point function in models with a modulation of the inflaton potential, where the scale dependence is not negligible. Finally, we generalize (some of) the consistency relations involving tensors and soft internal momenta.
}
\\
\noindent
\hrule

\allowdisplaybreaks


\section{Introduction and main results}
The initial conditions of the Universe are described by the correlation functions of the curvature perturbation $\zeta$ and of primordial tensor modes $\gamma$. These correlation functions are expected to be approximately scale-invariant, and indeed this matches what is observed for the 2-point function of $\zeta$. In many physical systems dilation invariance is promoted to full conformal invariance, so that it is natural to ask what is the role of conformal transformations for the statistics of primordial perturbations. In this paper we study this question focussing on (generic) single-field models.

First of all, let us understand why dilations and full conformal invariance have a different status in inflation. Inflation takes place in an approximate de Sitter space
\be
ds^2 = -dt^2 + e^{2 H t}  d \vec x^2 \;,
\label{eq:dS}
\ee
a maximally symmetric space whose isometry group is $SO(4,1)$, which coincides with the group of conformal transformations in 3-dimensional Euclidean space.  The time-evolving inflaton background is homogeneous and rotationally invariant, so that translations and rotations are exact symmetries of the whole system and show up---in an obvious way---at the level of correlation functions. This accounts for 6 out of the 10 generators of $SO(4,1)$. The dilation isometry
\be
\label{eq:dils}
t \to t - H^{-1}\log \lambda\;\qquad \vec x \to \lambda \vec x
\ee
is not, on general grounds, a symmetry of the inflaton background. Indeed the solution---which we can always assume to be of the form $\phi = t$ by a suitable field redefinition---is not invariant under a time shift. However it is quite common that there is an additional (approximate) symmetry, the shift symmetry $\phi \to \phi + c$. In this case the inflaton background is invariant under a combination of a dilation and a shift of the field\footnote{One can also consider the case in which the system is only invariant under a discrete (and not continuous) shift symmetry, see \cite{arXiv:1111.3373}.}. This unbroken symmetry is the origin of the approximate scale-invariance of the inflationary correlation functions \cite{arXiv:1007.0027}.   In Fourier space dilations act as 
\be
\zeta_{\vec k} \to \lambda^{-3} \zeta_{\vec k/\lambda} \;,
\ee 
so that the 2-point function is constrained to be of the form
\be
\langle \z_{\vec k_1} \z_{\vec k_2} \rangle = (2\pi)^3 \delta(\vec k_1 + \vec k_2) \frac{1}{k_1^3} F(k_1 e^{-H t}) \;.
\ee
Since $\zeta$ becomes time-independent when out of the Hubble radius, the function $F$ must become a constant in this limit. Therefore the spectrum must be of the scale-invariant form. The same argument implies that the $n$-point functions of $\zeta$ are product of the momentum-conserving $\delta$-function and a homogeneous function of degree $3(1-n)$.

Besides translations, rotations and dilations, de Sitter possesses three additional isometries 
\be
\label{eq:special}
t \to t - 2 H^{-1}  \vec b \cdot \vec x \;,\quad x^i \to x^i - b^i(-H^{-2} e^{-2 H t} + \vec x^2) +2 x^i (\vec b \cdot \vec x) \;,
\ee
parametrized by the infinitesimal vector $\vec b$. Notice that these symmetries at large times, $t \to +\infty$, act on the spatial coordinates as special conformal transformations. The reason why these symmetries are not so popular in the study of inflation is that, usually, correlation functions are not invariant since the inflaton background $\phi = t$ breaks these three symmetries.  On the other hand, these symmetries are explicit when the source of perturbations does not couple (or couples only slightly) with the inflaton: this is the case of tensor modes \cite{arXiv:1104.2846} and of scalar perturbations produced by a second field \cite{arXiv:1103.4164,Creminelli:2011mw}. One may wonder whether it is possible, like in the case of dilations, to impose an additional symmetry on the inflaton Lagrangian in such a way that some combination of the isometries \eqref{eq:special} and this internal symmetry is preserved. This would require that the inflaton action is invariant under 
\be
\label{eq:Gal}
\phi \to \phi + \vec b \cdot \vec x \;,
\ee
so that it is possible to cancel the variation induced on $\phi$ by the isometries \eqref{eq:special}. This is a Galilean transformation \cite{arXiv:0811.2197} with a spatial transformation parameter. It is certainly possible to write an inflaton action which is Galilean invariant \cite{arXiv:1009.2497,arXiv:1011.3004}, but it is well known that this symmetry is only well defined in the Minkowski limit and broken by a curved background \cite{arXiv:0901.1314}. This implies that correlation functions, which probe the theory at a scale comparable with the curvature, cannot be invariant under the isometries \eqref{eq:special} \footnote{In exact de Sitter and in the decoupling limit, interactions of the form $(\Box\phi)^n$ are invariant under spatial Galilean transformations. However, it is easy to check that all these couplings are redundant, i.e.~they can be set to zero by a proper field redefinition. This field redefinition is of the schematic form $\phi \to \phi + (\Box\phi)^{n-1}$ and thus vanishes on large scales, giving trivial cosmological correlation functions. Correlation functions are not trivial on short scales and are obviously $SO(4,1)$ invariant because they are produced by a de Sitter invariant field redefinition.}.

However, this is not the end for special conformal transformations in single-field inflation. Indeed, in the case of dilations we know that when we depart from exact shift symmetry (or even exact de Sitter geometry) the variation of the $n$-point function of $\zeta$ under dilation---which is now not zero---is related to the squeezed limit of the $(n+1)$-point function \cite{Maldacena:2002vr,Creminelli:2004yq,Cheung:2007sv}. The purpose of this paper is to generalize this argument to the special conformal transformations and to derive {\em conformal consistency relations} for single-field models.

We will start in Section \ref{sec:dS} to consider the simplest case, when the background metric can be approximated with de Sitter space and the inflaton perturbations can be studied in the decoupling limit, neglecting the effect on the metric. Moreover, we will assume exact scale invariance, i.e.~exact shift symmetry on $\phi$. Within these approximations the metric is de Sitter and thus invariant under \eqref{eq:special}. Special conformal transformations are non-linearly realized, since eq.~\eqref{eq:Gal} induces a linear dependence on $\vec x$ of the background solution (similarly, dilations induce a homogeneous shift, $\phi \to \phi + c$, of the background). This implies that, in the same way the effect of a homogeneous long-wavelength mode on an $n$-point function is given by a dilation, the effect of a constant gradient is given by a special conformal transformation. 
We thus obtain a relation between the variation of an $n$-point function under a special conformal transformation and the squeezed limit of the $(n+1)$-point function, in particular with the term going as $1/q^2$ as the momentum $q$ goes to zero. This generalizes the consistency relations for dilations, which relate the variation of an $n$-point function under dilations with the $1/q^3$ divergence of the $(n+1)$-point function.  
As the conformal variation of the 2-point function trivially vanishes, we recover the known result that the 3-point function does not contain terms going as $1/q^2$ in the squeezed limit \cite{Creminelli:2011rh}. The first non-trivial relation is between the 3- and the 4-point function and we verify it for models with a reduced speed of sound. In this case these consistency relations capture, at the level of observables, the relation among operators induced by the non-linear realization of Lorentz invariance in the Lagrangian.

In Section \ref{sec:gravity} and Appendix \ref{app:consaction} the conformal consistency relations are generalized beyond the decoupling limit, i.e.~when perturbations of the metric become relevant. Even though in this case the metric deviates from de Sitter, and therefore does not possess the $SO(4,1)$ isometry, the conformal consistency conditions can still be derived, using the fact that a special conformal transformation of the spatial coordinates induces a profile of $\zeta$ which is linear in $\vec x$. In other words, with a suitable change of coordinates which acts on the spatial ones as a conformal transformation, one can add   to a given solution a long-wavelength $\zeta$ mode including the first gradient corrections. This is similar in spirit to the treatment of adiabatic modes done in \cite{Weinberg:2003sw}; we study an example of this technique (unrelated to inflation) in Appendix \ref{app:MD}. In this way the conformal consistency relations can be applied to single-field slow-roll inflation and in fact we are able to check the relation between the 3- and 4-point functions for this class of models. 

In Section  \ref{sec:tilt} we generalize our consistency relations to include deviations from scale-invariance, in which the squeezed limit contains both terms going as $1/q^3$ and terms going as $1/q^2$. As an example we verify the obtained consistency relations for all the $n$-point functions in models with resonant non-Gaussianity, where the deviation from scale-invariance cannot be neglected. The impatient reader can look at the most general consistency relation, eq.~\eqref{eq:mastertilt}.
In Section \ref{sec:tensors} and Appendix \ref{app:GW} we generalize some of the consistency relations involving tensor modes and we explain why this is only possible in some cases. Finally, in Section \ref{sec:internal} we generalize the consistency relations for the case in which one of the internal momenta is small. Conclusions and future prospects are discussed in Section \ref{sec:conclusions}.

\section{\label{sec:dS}The limit of exact de Sitter}

In this Section we consider single-field inflation, in the limit in which the background metric can be approximated with de Sitter space, i.e.~at zeroth order in slow-roll. 
Furthermore, in most inflationary models one can study the inflaton perturbations as decoupled from gravity and thus take the metric as unperturbed. In this Section we assume that this is the case, and work in this decoupling limit. We also assume exact scale invariance, i.e.~exact shift symmetry on $\phi$. When these approximations hold, the metric is given by an unperturbed de~Sitter metric, equation \eqref{eq:dS}, and is invariant under the full isometry group $SO(4,1)$. 

Without loss of generality, through a field redefinition, one can take the background solution for the inflaton field to be $\phi_0(t) = t$. Let us parametrize the perturbations around this background as
\be
\phi(t,\vec{x}) \equiv t + \pi(t,\vec{x})\,.
\ee
The background evolution of the inflaton breaks some of the isometries of de Sitter: dilations and special conformal transformations (eqs.~\eqref{eq:dils} and \eqref{eq:special}). We are interested in studying the conditions that the non-linear realization of these isometries imposes on the $n$-point correlation functions of the field perturbations.

\subsection{Derivation of the consistency relations in exact de Sitter}

As a warm-up let us consider dilations first. It is easy to see that under a dilation parametrized by $\lambda$, eq.~\eqref{eq:dils}, the perturbations transform as
\be
\pi(x) \rightarrow \tilde\pi(\tilde x) = \pi(x) + H^{-1}\log \lambda\,.
\label{eq:piDilation}
\ee
Since the action is here assumed to be invariant under a shift of the field $\pi \rightarrow \pi + c$, the background is invariant under a diagonal composition of a dilation and a shift. Thus, the $n$-point correlation functions are expected to be invariant under this diagonal transformation\footnote{Note that this is indeed the diagonal transformation; in transforming the $n$-point correlation function we cancel the shift of the field induced by the dilation with an opposite shift. In other words the diagonal transformation amounts just to a rescaling of the coordinates in the $\pi$ correlation functions.}
\be
\delta_d\langle \pi_{\vec{k}_1} \dots \pi_{\vec{k}_n} \rangle \equiv (2\pi)^3 \delta(\vec{k}_1+\dots+\vec{k}_n) \bigg[3(n-1) + \sum_{a = 1}^{n} \vec{k}_a\cdot \vec \partial_{k_a}\bigg]\langle \pi_{\vec{k}_1} \dots \pi_{\vec{k}_n} \rangle' = 0\,,
\label{eq:piCorrelationDilation}
\ee
where $\langle\dots\rangle'$ denotes, here and in the following, a correlation function where the Dirac delta $(2\pi)^3 \delta(\vec{k}_1+\dots+\vec{k}_n) $ of momentum conservation has been dropped. This implies that the correlation functions in Fourier space for large $t$ are homogeneous functions of the momenta of degree $-3(n-1)$, that is
%
%
%
\be
\langle \pi_{\alpha \vec{k}_1} \dots \pi_{\alpha \vec{k}_n} \rangle' =  \frac{1}{\alpha^{3(n-1)}} \langle \pi_{\vec{k}_1} \dots \pi_{\vec{k}_n} \rangle' \;.
\ee

The unbroken diagonal symmetry also has another interesting and well-known consequence, namely the vanishing of the squeezed limit of a $n$-point function. In order to see this, take the long-wavelength mode $\pi_L$ to be approximately space independent\footnote{We are also assuming $\pi_L$ to be constant in time. This is true out of the horizon in the class of models under consideration in this Section up to slow-roll corrections.}. Note that equation \eqref{eq:piDilation} implies that we can approximate the effect of such a constant long-wavelength mode on an $n$-point function as a rescaling of the coordinates
\be
\langle \pi(\vec{x}_1) \dots \pi(\vec{x}_{n}) \rangle_{\pi_L} = \langle\pi(\lambda\vec{x}_1) \dots \pi(\lambda\vec{x}_{n}) \rangle\,,
\ee
where $\lambda \equiv e^{-H\pi_L}$. However, equation \eqref{eq:piCorrelationDilation} tells us that a rescaling of coordinates leaves the $n$-point function invariant. There is then no correlation between an $n$-point function and a long-wavelength mode: the squeezed limit of the $(n+1)$-point function vanishes. 

\begin{figure}[!!!h]
\begin{center}
\includegraphics[width=0.9 \textwidth]{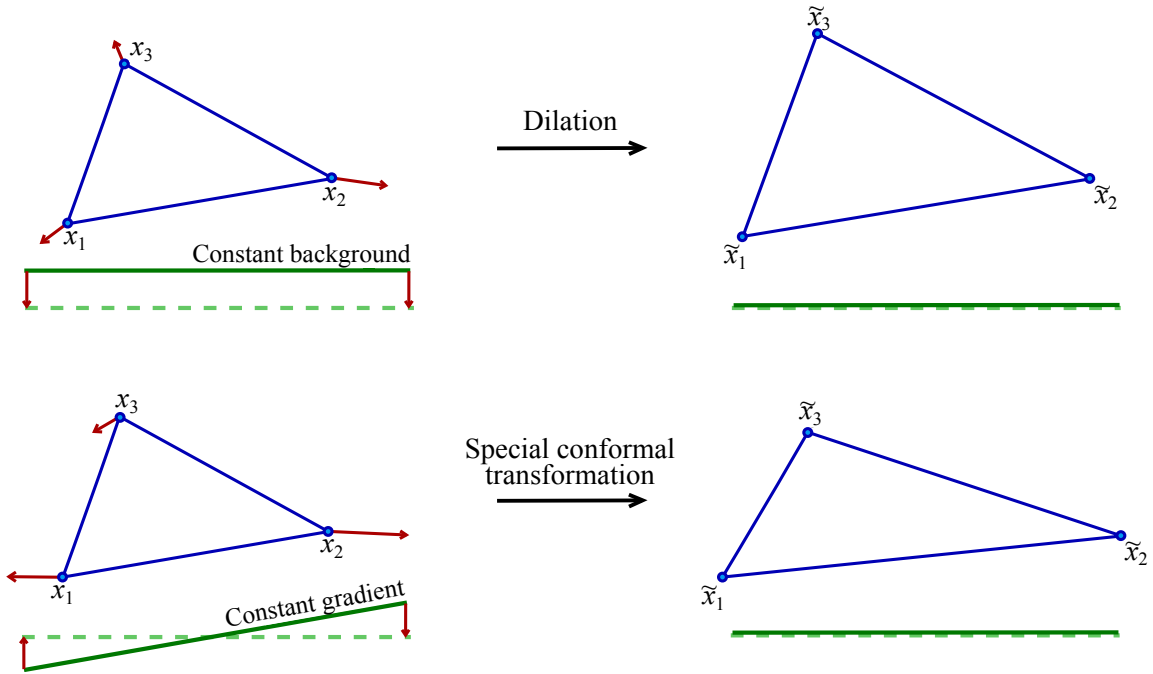}  
\end{center}
\caption{\small {\em A constant background mode can be removed by a dilation. A background with a constant gradient can be removed by a special conformal transformation.}}
\label{fig:dilconf}
\end{figure}

If we were to drop the assumption of the shift symmetry, the variation of the $n$-point function under a rescaling of the coordinates would be different from zero. The argument above tells us that in that case the squeezed limit of the $(n+1)$-point function would be proportional to the deviation of the $n$-point function from scale invariance (see figure \ref{fig:dilconf})
\be
\langle \pi_{\vec{q}} \pi_{\vec{k}_1} \dots \pi_{\vec{k}_{n}} \rangle' \overset{q\rightarrow 0}{=} H P_\pi(q) \bigg[3(n-1) + \sum_{a = 1}^{n} \vec{k}_a\cdot \vec \partial_{k_a}\bigg]\langle \pi_{\vec{k}_1} \dots \pi_{\vec{k}_{n}} \rangle'\,,
\ee
where $P_\pi$ is the power spectrum of the field perturbations defined as
\be
\langle \pi_{\vec{k}_1}\pi_{\vec{k}_2}\rangle = (2\pi)^3\delta(\vec{k}_1 + \vec{k}_2)P_\pi(k_1)\,.
\ee
The reader can consult Reference \cite{Cheung:2007sv} or Section \ref{sec:tilt} for a precise derivation of this result. We thus obtain a relation between the squeezed limit of the $(n+1)$-point function and the dilation of the $n$-point function, given by the non-linearly realized dilation symmetry. We will now see that the same happens when considering the special conformal transformations which are spontaneously broken.

Let us start by expanding the long-wavelength mode to linear order in $\vec{x}$ thus\footnote{Without loss of generality we are expanding around $\vec{x} = 0$. We will see in Section \ref{sec:tilt} that our results do not depend on this choice.} 
\be
\pi_L(x) = \pi_L(0) + \partial_i\pi_L(0) x^i.
\label{eq:piExpansion}
\ee
We saw above that the constant piece $\pi_L(0)$ can be induced by a rescaling of coordinates which, in the presence of a shift symmetry, leaves the correlation functions invariant. We will now focus on the linear piece $\partial_i\pi_L(0)x^i$.
Let us write down the transformation of the perturbations under a special conformal transformation, eq. \eqref{eq:special}, parametrized by the infinitesimal parameter $\vec{b}$
\be
\pi(x) \rightarrow \tilde\pi(\tilde x) = \pi(x) + 2 H^{-1} \vec{b}\cdot\vec{x}\,.
\label{eq:piSpecial}
\ee
According to this relation a constant gradient can be induced by a special conformal transformation on the coordinates with $\vec{b} = -(1/2) H \vec{\partial}\pi_L(0)$ \; (\footnote{The long wavelength mode is taken to be time-independent: we are implicitly assuming that its time-dependence arises at ${\cal O}(k^2)$ and can thus be neglected for our purposes.})
\be
\langle \pi(\vec{x}_1) \dots \pi(\vec{x}_{n}) \rangle_{\partial\pi_L} = \langle\pi(\vec{\tilde x}_1) \dots \pi(\vec{\tilde x}_{n}) \rangle\,,
\ee
where $\tilde{x}^i \equiv x^i - b^i\vec x^2 +2 x^i (\vec b \cdot \vec x)$. This equation holds in the out-of-the-horizon limit, when one can drop the time-dependent piece in the transformation of the $x^i$ in the isometry \eqref{eq:special} and neglect the time-dependence of $\pi$. As in the case of the broken scale invariance, the variation of the correlation function under such a coordinate transformation is not zero and will therefore induce a relation between the variation of the $n$-point function and the squeezed limit of the $(n+1)$-point function (see figure \ref{fig:dilconf}). Let us proceed by computing this variation in Fourier space \cite{arXiv:1104.2846}
\begin{align}
\delta_b \langle\pi(\vec{\tilde x}_1) \dots \pi(\vec{\tilde x}_{n}) \rangle &= \int \frac{\mathrm{d}^3 \vec k_1}{(2\pi)^3}\dots\frac{\mathrm{d}^3 \vec k_{n}}{(2\pi)^3} \langle\pi_{\vec{k}_1} \dots \pi_{\vec{k}_{n}} \rangle \nonumber \\ 
	&\phantom{=\int} \times (-i) \sum_{a=1}^{n} k_a^i\Big[b^i \vec x_a^2 -2 x_a^i (\vec b \cdot \vec x_a)\Big] e^{i(\vec{k}_1\cdot\vec{x}_1 + \dots + \vec{k}_{n}\cdot\vec{x}_{n})}  \nonumber\\
&= \int \frac{\mathrm{d}^3 \vec k_1}{(2\pi)^3}\dots\frac{\mathrm{d}^3 \vec k_{n}}{(2\pi)^3} e^{i(\vec{k}_1\cdot\vec{x}_1 + \dots + \vec{k}_{n}\cdot\vec{x}_{n})} \nonumber \\
&\phantom{=\int} \times (-i) \sum_{a=1}^{n} b^i\Big( 6\partial^i_{k_a} - k^i_a \vec{\partial}_{k_a}^2 +2\vec{k}_a \cdot \vec{\partial}_{k_a}  \partial^i_{k_a} \Big)\langle\pi_{\vec{k}_1} \dots \pi_{\vec{k}_{n}} \rangle\,.
\end{align}
We can thus compute the $(n+1)$-point correlation function where one of the modes is much longer than the other $n$ to linear order in the gradient of the long mode
\begin{align}
\langle \pi_{\vec{q}}\pi_{\vec{k}_1}\dots\pi_{\vec{k}_{n}}\rangle &\overset{q\rightarrow 0}{=} \big\langle\pi_{\vec{q}}\langle\pi_{\vec{k}_1}\dots\pi_{\vec{k}_{n}}\rangle_{\partial\pi_L}\big\rangle \nonumber \\
& = (-i) \langle\pi_{\vec{q}} \, b^i\rangle' \sum_{a=1}^{n} \Big( 6\partial^i_{k_a} - k^i_a \vec{\partial}_{k_a}^2 +2\vec{k}_a \cdot \vec{\partial}_{k_a}  \partial^i_{k_a} \Big)\langle\pi_{\vec{k}_1} \dots \pi_{\vec{k}_{n}} \rangle \nonumber \\
& = \frac{H}{2} P_{\pi}(q)q^i\sum_{a=1}^{n} \Big( 6\partial^i_{k_a} - k^i_a \vec{\partial}_{k_a}^2 +2\vec{k}_a \cdot \vec{\partial}_{k_a}  \partial^i_{k_a} \Big)\langle\pi_{\vec{k}_1} \dots \pi_{\vec{k}_{n}} \rangle\,,
\end{align}
where in the second equality we used the fact that a constant mode leaves the correlation functions unaffected in the presence of the field shift symmetry. Since within the approximations used in this Section one can relate the curvature perturbation in comoving gauge $\zeta$ with the field perturbation by $\zeta = -H\pi$, we rewrite this last equation as\footnote{Here we used the fact that $\delta(\vec{q} + \vec{k}_1 + \dots + \vec{k}_{n}) \approx \delta(\vec{k}_1 + \dots + \vec{k}_{n})$ up to subleading corrections in $q$, and that the transformation commutes with the Dirac delta (see Section \ref{sec:tilt}).}
\be
\label{eq:CCRdS}
\langle \zeta_{\vec{q}}\zeta_{\vec{k}_1}\dots\zeta_{\vec{k}_{n}}\rangle' \overset{q\rightarrow 0}{=} -\frac{1}{2} P(q)q^i\sum_{a=1}^{n} \Big( 6\partial^i_{k_a} - k^i_a \vec{\partial}_{k_a}^2 +2\vec{k}_a \cdot \vec{\partial}_{k_a}  \partial^i_{k_a} \Big)\langle\zeta_{\vec{k}_1} \dots \zeta_{\vec{k}_{n}} \rangle'\,,
\ee
where $P$ is the power-spectrum of $\zeta$. 

In the case of the $2-$ and $3-$point functions, the correlator only depends on the magnitudes of the momenta $|\vec{k}_a|=k_a$; in this case the conformal transformation simplifies to \cite{arXiv:1104.2846}
\begin{equation}\label{eq:confsimple}
\sum_{a=1}^{n} \vec{q}\cdot \vec{k}_a \left[ \frac{4}{k_a}\frac{\partial}{\partial k_a} + \frac{\partial^2}{\partial k_a^2} \right] \;.
\end{equation}
In this form it is easy to see that the conformal variation of the $2-$point function vanishes (up to terms which are singular in Fourier space, see \cite{arXiv:1104.2846} and Section \ref{sec:tilt}), as the two terms in the sum cancel. This is obvious from the fact that the 2-point function does not specify any direction with which the vector $\vec b$ can contract. This implies that the improved consistency relation just implies that the 3-point function is suppressed, compared with a local shape, by (at least) ${\cal O}(q/k)^2$  \cite{Creminelli:2011rh}. Much less trivial is the implication of eq.~\eqref{eq:CCRdS} for the squeezed limit of the 4-point function, as we are now going to discuss.

\subsection{Small speed of sound}
In this Section we are going to check the conformal consistency relation between the 3- and 4-point functions in inflationary models with reduced speed of sound. These are described by a Lagrangian of the form
\be
S=\frac{1}{2}\int \rmd^4x\sqrt{-g}\left( M_P^2 R+2P(X,\phi) \right) \;,
\ee
where $X=-\frac{1}{2}\partial_\mu\phi\partial^\mu\phi$. In order to compute the 3- and 4-point functions we have to expand the action up to fourth order, find the interaction Hamiltonian and use the standard in-in formalism. All these computations have been done in \cite{Chen:2009bc,Arroja:2009pd} and here we follow the notation of the first paper. Any $n$-point function can be written as
\be
\langle \zeta_{\vec{k}_1} \cdots \zeta_{\vec{k}_n} \rangle = (2\pi)^3\delta(\vec{k}_1+\cdots +\vec{k}_n) P_\zeta^{n-1} \prod_{i=1}^n\frac{1}{k_i^3} \mathcal{M}^{(n)}(\vec{k}_1,...,\vec{k}_n)
\ee
and the 2-point function is given by
\be
\langle \zeta_{\vec{k}_1} \zeta_{\vec{k}_2} \rangle = (2\pi)^3\delta(\vec{k}_1+\vec{k}_2) P_\zeta \frac{1}{2k_1^3}, \quad P_\zeta=\frac{1}{2\Mpl^2}\frac{H^2}{c_s\epsilon} \;,
\ee
with $\epsilon \equiv - \dot H/H^2$ and the speed of sound defined as
\be
c_s^2 \equiv \frac{P_{,X}}{P_{,X}+2XP_{,XX}} \;.
\ee

The amplitude of the 3-point function is:
\be
\mathcal{M}^{(3)}=\left(\frac{1}{c_s^2}-1-\frac{2\lambda}{\Sigma}\right)\frac{3k_1^2k_2^2k_3^2}{2k_t^3} + \left( \frac{1}{c_s^2}-1 \right) \left( -\frac{1}{k_t}\sum_{i>j} k_i^2k_j^2 +\frac{1}{2k_t^2}\sum_{i\neq j}k_i^2k_j^3 +\frac{1}{8}\sum_i k_i^3  \right) \;,
\ee
where $k_t = k_1 + k_2+ k_3$ and the parameters $\lambda$ and $\Sigma$ are related to derivatives of the Lagrangian with respect to $X$:
\begin{align}
& \lambda = X^2P_{,XX}+\frac{2}{3}X^3P_{,XXX} \nonumber \\
& \Sigma = X^2P_{,X}+2X^2P_{,XX} \;.
\end{align}

The 4-point function contains two kind of contributions. One comes from diagrams with a quartic interaction and the other from scalar-exchange diagrams. It is easy to realize that scalar-exchange diagrams do not contribute to the conformal consistency relation since they vanish as $k_4^2$ for $k_4 \to 0$. This is obvious when a time derivative acts on the soft external leg with momentum $k_4$, as the time derivative of the wavefunction gives $k_4^2$. The same holds when a spatial derivative acts on the soft external leg. Indeed at the relevant vertex we will have external momenta $\vec{k}_4$ and (say) $\vec{k}_3$ and an internal momentum $- \vec k_4 - \vec k_3$. Now, depending on which leg the second spatial derivative acts we have two diagrams proportional to $\vec{k}_4\cdot\vec{k}_3$ and $-\vec{k}_4\cdot (\vec{k}_4+\vec{k}_3)$, which cancel each other at leading order, leaving terms of order $k_4^2$ (\footnote{Notice that it is crucial, for this argument to work, that time derivatives in the propagator are inside the T-ordering (or anti T-ordering), so that they simply act on the wavefunctions. In the standard S-matrix calculation in Minkowski, one usually has the time derivatives acting outside the $\theta$ functions to restore Lorentz invariance, with the contact terms canceling the Hamiltonian terms which are not Lorentz invariant.}). This behaviour can be checked in the explicit results of \cite{Chen:2009bc,Arroja:2009pd}

We can therefore concentrate on the diagrams with a quartic interaction, whose contribution to the amplitude is given by \cite{Chen:2009bc}
\begin{align}
\mathcal{M}&^{(4)}_{cont}  = \left[ \frac{3}{2}\left( \frac{\mu}{\Sigma}-\frac{9\lambda^2}{\Sigma^2} \right) \frac{\prod_{i=1}^4k_i^2}{k_t^5} - \frac{1}{8}\left( \frac{3\lambda}{\Sigma}-\frac{1}{c_s^2} +1 \right)\frac{k_1^2k_2^2(\vec{k}_3\cdot\vec{k}_4)}{k_t^3}\left( 1+\frac{3(k_3+k_4)}{k_t} +\frac{12k_3k_4}{k_t^2} \right) \right. \nonumber \\
& \left. + \frac{1}{32} \left( \frac{1}{c_s^2} - 1 \right) \frac{(\vec{k}_1\cdot\vec{k}_2)(\vec{k}_3\cdot\vec{k}_4)}{k_t} \left( 1+\frac{\sum_{i<j}k_ik_j}{k_t^2} +\frac{3k_1k_2k_3k_4}{k_t^3}\sum_{i=1}^4\frac{1}{k_i} +\frac{12k_1k_2k_3k_4}{k_t^4} \right) \right] + \mbox{23 perm.}
\end{align}
with $\mu$ defined as:
\be
\mu=\frac{1}{2}X^2P_{,XX} + 2X^3P_{,XXX}+\frac{2}{3}X^4 P_{,XXXX} \;.
\ee
We are interested in terms linear in $k_4$ which are given by 
\begin{align}
\mathcal{M}&^{(4)}_{cont}|_{\vec{k}_4\rightarrow 0}  = \left[  - \frac{1}{2}\left( \frac{3\lambda}{\Sigma}-\frac{1}{c_s^2} +1 \right)\frac{k_1^2k_2^2(\vec{k}_3\cdot\vec{k}_4)}{k_t^3}\left( 1+\frac{3k_3}{k_t}  \right) \right. \nonumber \\
& \left. + \frac{1}{4} \left( \frac{1}{c_s^2} - 1 \right) \frac{(\vec{k}_1\cdot\vec{k}_2)(\vec{k}_3\cdot\vec{k}_4)}{k_t} \left( 1+\frac{k_1k_2+ k_1k_3+ k_2k_3}{k_t^2} +\frac{3k_1k_2k_3}{k_t^3}  \right) \right] + \mbox{2 terms} \;,
\end{align}
where the two additional terms have the same form with $\vec{k}_1\leftrightarrow\vec{k}_3$ and $\vec{k}_2\leftrightarrow\vec{k}_3$. Notice that since in the squeezed limit $\vec{k}_1$, $\vec{k}_2$ and $\vec{k}_3$ form a triangle, we can replace $(\vec{k}_1\cdot\vec{k}_2)\rightarrow \frac{1}{2}(k_3^2-k_1^2-k_2^2)$. Factoring out $(\vec k_3 \cdot \vec k_4)$, we can see that the expression has the same structure as the right-hand side of the conformal consistency conditon:
\be
\sum_{i=1}^3(\vec{k}_4\cdot\vec{k}_i)\left( \frac{4}{k_i}\frac{\partial}{\partial k_i} + \frac{\partial^2}{\partial k_i^2} \right)\langle \zeta_{\vec{k}_1} \zeta_{\vec{k}_2} \zeta_{\vec{k}_3}\rangle \;.
\ee
Indeed it is straightforward to check that eq.~\eqref{eq:CCRdS} is satisfied. (To simplify the calculations one can choose, without loss of generality, $\vec{k}_4=(1,0,0)$. Notice that one has to impose $\vec{k}_1+\vec{k}_2 +\vec{k}_3=0$ at the end, after taking all the derivatives.)

Let us comment on the information carried by the conformal consistency conditions. 

\begin{itemize}

\item At the level of the Lagrangian, it is well known that operators with a different number of fields are related due to the non-linear realization of the Lorentz symmetry \cite{Cheung:2007st}. The conformal consistency relations contain the same information at the level of observable correlation functions. For example, for $c_s =1$, the same operator proportional to $\lambda$ (it is the operator $(g^{00}+1)^3$ in the language of the effective field theory of inflation \cite{Cheung:2007st}), contributes both to the 3-point function and to the 4-point function. This is encoded in the conformal consistency relation, as the conformal variation of the 3-point function must be matched by a corresponding (squeezed limit of the) 4-point function.

\item Similarly, if an operator only contributes to the 4-point function and not to the 3-point one, then this 4-point function must have a suppressed squeezed limit, i.e.~the coefficient of the $q^{-2}$ divergence must vanish. This is indeed what happens for the operator proportional to $\mu$ above ($(g^{00}+1)^4$ in the language of the effective field theory of inflation \cite{Cheung:2007st}). In \cite{Senatore:2010jy} it was shown that it is technically natural to make this operator large and the 4-point function arbitrarily large compared to the 3-point function: in any case this happens for any $n$-point function, its $q^{-2}$ divergence must vanish. 

\item The conformal consistency relation eq.~\eqref{eq:CCRdS} states that the $q^{-2}$ divergence of the 4-point function is {\em always small}, of order $P_\zeta^{1/2}$ times the 3-point function. This is not at all obvious at the level of the action. Indeed, the operator which reduces the speed of sound gives a 4-point function of order $P_\zeta^{1/2} c_s^{-2}$ compared with the 3-point function. Moreover, this operator is of the form $(\nabla\pi)^4$ so that it naively gives rise to a $q^{-2}$ squeezed limit. It turns out that, in going to the Hamiltonian, the term $(\nabla\pi)^4$ gets a $c_s^2$ suppression while the leading source of the 4-point function ($\propto c_s^{-4}$) is given by the exchange diagrams which, as discussed above, do not contribute to the $q^{-2}$ divergence. The same kind of cancellation occurs for the operator $\lambda \dot\pi^2 (\nabla\pi)^2$ which, in the small $c_s$ limit, would naively give a 4-point which is too large in the squeezed limit to satisfy eq.~\eqref{eq:CCRdS}. Again, this term is cancelled in going to the Hamiltonian and gets suppressed by $c_s^2$, while the leading 4-point function is given by the exchange diagrams.

\item Equation \eqref{eq:CCRdS} encodes the non-linear realization of the de Sitter isometries; indeed it looks suggestively similar to a Ward identity for non-linearly realized symmetries. Notice, however, that cosmologically we only observe correlation functions at late times, so that we unavoidably miss some information. For example, the late time power spectrum does not distinguish between the de Sitter invariant case and the case with $c_s<1$ when de Sitter isometries are non-linearly realized.

\end{itemize}

\section{\label{sec:gravity}Including gravity}
We are now going to see that the conformal consistency relations, which were derived above from the non-linear realization of the de Sitter isometry group $SO(4,1)$, hold much more generally, basically for any single-field model. In this more general case, the group $SO(4,1)$ arises as the group of conformal transformation of 3d Euclidean space. First of all, we are going to show that a long wavelength mode can be (locally) created by a suitable change of coordinates, even including the first gradient corrections.
\subsection{Adiabatic modes at order $k$}
We want to show how to construct an adiabatic solution in $\zeta$-gauge, which is correct not only in the homogeneous limit, but also including corrections linear in the gradients. An adiabatic mode is locally a coordinate redefinition of the unperturbed solution: we thus parallel what Weinberg \cite{Weinberg:2003sw} does in Newtonian gauge and look for the most general transformation of the unperturbed FRW that leaves the metric in $\zeta$-gauge. As the inflaton is unperturbed in this gauge, the slicing is fixed and we cannot redefine the time variable. Since we are not interested in tensor modes, we want our change of coordinates to induce only a conformal rescaling of the spatial metric, in this way ``exciting'' only $\zeta$. The most general transformation we can do is thus an element of the 3d conformal group $SO(4,1)$, a different one at each time $t$. This transformation will turn on $\zeta$, that will be given---at each $t$---by a term linear in $\vec x$ and a constant piece: 
\be
\zeta = 2 \vec b(t) \cdot \vec x + \lambda(t) \;.
\ee 
Now we have to check what are the additional constraints coming from the requirement that the solutions we found are the $k \to 0$ limit of physical solutions \cite{Weinberg:2003sw}. For standard slow-roll inflation the linear momentum constraint reads
\be
\label{eq:linearmomentum}
\partial_j(H \delta N - \dot \zeta) = 0 \;,
\ee
while the Hamiltonian constraint reads
\be
\label{eq:linearH}
(3 H^2 + \dot H) \delta N  + H \partial_i N^i= - \frac{\nabla^2}{a^2}\zeta + 3 H \dot\zeta \;.
\ee
The coordinate transformations we discussed do not induce $\delta N$ (at linear order). Therefore, from the first equation, we have to require $\dot\zeta = 0$ up to order $k^2$ if we want our coordinate transformation to describe an adiabatic solution at order $k$. This implies that $\vec b$ and $\lambda$ must be taken as time-independent
\be
\label{eq:zetalong}
\zeta = 2 \vec b \cdot \vec x + \lambda \;.
\ee
The Hamiltonian constraint implies that $N^i$ is of order $k$. Its spatial dependence would be of order $k^2$ and can thus be neglected within our approximation: $N^i(t)$ just depends on time. We can always generate this $N^i(t)$ by a time-dependent translation of the spatial coordinates 
\be
\tilde x^i = x^i + \delta x^i(t) \quad\to\quad N^i(t) = - \frac{d}{d t}\delta x^i(t) \;.
\ee
This does not change the metric at constant $t$, i.e.~it does not change $\zeta$. The situation is very similar for a general model of single-field inflation. In this case the constraint equations are different, but from their general form (see for example \cite{Creminelli:2011rh}) one can still conclude that $\dot\zeta$ is of order $k^2$ and $N^i$ of order $k$, which is all that we need in the discussion above.

Notice that in solving the constraints \eqref{eq:linearmomentum} and \eqref{eq:linearH} we have to choose boundary conditions. We choose them in a standard way, for example taking $\zeta$ to vanish at infinity: this does not clash with eq.~\eqref{eq:zetalong}, as this is only a local approximation of the solution which, once the constraints are solved, can be extended to a regular global solution.

In conclusion, it is always possible to write an adiabatic solution which is valid including ${\cal O}(k)$ corrections: this is obtained by a (time-independent) special conformal transformation and dilation of the spatial coordinates and a time-dependent translation.

\subsection{Derivation of the conformal consistency relations including gravity}
We showed that a long-wavelength mode can be obtained---up to correction of order $k^2$---as a coordinate transformation from the unperturbed FRW. This property holds also in the presence of short modes. Indeed, in Appendix \ref{app:consaction} we explicitly check, in the case of minimal slow-roll inflation, that the cubic action can be obtained with a coordinate transformation of the quadratic action, up to corrections which are quadratic in the long-mode momentum.
In general, one can constrain the behavior of any $(n+1)$-point function in the limit in which one of the momenta $q$ goes to zero.  
Indeed, the effect of this long-wavelength mode on the remaining $n$-point function can be obtained from the $n$-point function in the absence of the background mode, simply by a coordinate transformation, up to corrections of order $q^2$. The coordinate transformation acts on surfaces of constant $t$ as the composition of a dilation, a special conformal transformation and a (time-dependent) translation. Given that the $n$-point function is translationally invariant and we are for the moment neglecting the (small) scale-dependence, the effect of the coordinate transformation simply reduces to the special conformal transformation.  

We can now repeat the arguments that we followed above in the decoupling limit. 
The infinitesimal variation of the $n$-point function under a conformal transformation with parameter $\vec b$ is given by
\begin{equation}
\delta\langle \zeta_{\vec{k}_1} \zeta_{\vec{k}_2} \ldots \zeta_{\vec{k}_{n}} \rangle '  = - i \sum_{a=1}^{n} \Big[ 6\vec{b}\cdot \vec{\partial}_{k_a} - \vec{b}\cdot \vec{k}_a \vec{\partial}_{k_a}^2 +2\vec{k}_a \cdot \vec{\partial}_{k_a} (\vec{b}\cdot \vec{\partial}_{k_a}) \Big] \langle \zeta_{\vec{k}_1} \zeta_{\vec{k}_2} \ldots \zeta_{\vec{k}_{n}} \rangle '   \;.
\end{equation}
This implies, using eq.~\eqref{eq:zetalong}, that the formula above with $\vec b = \vec\nabla \zeta /2$ gives the effect of the long mode on the $n$-point function. To get the $(n+1)$-point function, we have to multiply by the long $\zeta$, go to Fourier space for it, and finally average over this long mode, exactly as we did above. We get 
\be
\begin{split}
\label{eq:masterzeta}
\langle \zeta_{\vec q} \;\zeta_{\vec{k}_1} \zeta_{\vec{k}_2} \ldots \zeta_{\vec{k}_{n}}\rangle'  & \overset{q\rightarrow 0}{=}  -\frac12 P(q) q^i D_i  \langle \zeta_{\vec{k}_1} \zeta_{\vec{k}_2} \ldots \zeta_{\vec{k}_{n}}\rangle' + {\cal{O}}(q/k)^2 \\ 
\quad {\rm with } \quad q^iD_i & \equiv \sum_{a=1}^{n} \Big[ 6\vec{q}\cdot \vec{\partial}_{k_a} - \vec{q}\cdot \vec{k}_a \vec{\partial}_{k_a}^2 +2\vec{k}_a \cdot \vec{\partial}_{k_a} (\vec{q}\cdot \vec{\partial}_{k_a}) \Big] \;,
\end{split}
\ee
where $P(q)$ is the power spectrum at momentum $q$. 

This is exactly the same as eq.~\eqref{eq:CCRdS} that we derived above using the de Sitter isometries. Now we see that it holds in a much more general situation, even when the geometry is not close at all to the de Sitter geometry. Notice that in this more general case, the $SO(4,1)$ group has a different interpretation: the conformal group in 3 dimensions. However it is not too surprising that in the decoupling limit, when the geometry can be taken as non-dynamical, the equation can be derived in a way which has nothing to do with $\zeta$, which describes a metric perturbation, but only with $\pi$. Indeed in the decoupling limit, the $SO(4,1)$ appears as a symmetry of the action, non-linearly realized on $\pi$.

As discussed in the previous Section, this relation implies the absence of $q^{-2}$ terms in the squeezed limit of the 3-point function. In the next Subsection we use the results above to study the squeezed limit of the 4-point function for standard slow-roll inflation.

\subsection{Conformal consistency relation for slow-roll inflation}
%

Let us verify the conformal consistency relation \eqref{eq:masterzeta} for $n=4$, i.e.~relating the squeezed limit of the 4-point function with the conformal variation of the 3-point function, for slow-roll inflation with a standard kinetic term. The 3-point function is the classic Maldacena's result \cite{Maldacena:2002vr}
\begin{equation}\label{eq:Malda}
\langle \zeta_{\vec{k}_1} \zeta_{\vec{k}_2} \zeta_{\vec{k}_3} \rangle' =  \frac{H^4}{4\epsilon^2 M_P^4} \frac{1}{\prod (2k_i^3)} \left[ (2\eta-3\epsilon)\sum_{i}k_i^3 + \epsilon\sum_{i\neq j}k_i k_j^2 + \epsilon\frac{8}{k_t}\sum_{i>j} k_i^2 k_j^2 \right],
\end{equation}
where $k_t = k_1 + k_2+ k_3$ and $\eta$ is the second slow-roll parameter, related to the inflaton potential $V$ by $\eta \equiv M_P^2 V''/V$. 

The 4-point function is the sum of two comparable contributions, one coming from 4-legs interactions of $\zeta$ \cite{Seery:2006vu}
and one from a graviton exchange \cite{Seery:2008ax}. The contribution from the graviton exchange will not enter in our conformal consistency relation as it decays too quickly in the squeezed limit $q \to 0$. This is easy to see from the cubic vertex that enters (twice) in the diagram:
\begin{equation}
S\sim\int d^4x \partial_i\zeta\partial_j\zeta\gamma_{ij} \;.
\end{equation}
As the graviton is transverse, we can rewrite this vertex moving the two derivatives on the leg with momentum $q$. This implies that the graviton exchange gives a contribution suppressed by two powers of $q$ in the squeezed limit compared with a local shape and this does not contribute to eq.~\eqref{eq:masterzeta}.

The contact term contribution is given by \cite{Seery:2006vu}:
\begin{equation}
\label{eq:4pfsr}
\langle \zeta_{\vec{k}_1} \zeta_{\vec{k}_2} \zeta_{\vec{k}_3} \zeta_{\vec{k}_4} \rangle'_{\rm cont} =  \frac{H^6}{4\epsilon^3 M_P^6} \frac{1}{\prod (2k_i^3)} \epsilon \sum_{perm} \mathcal{M}_4(\vec{k}_1,\vec{k}_2,\vec{k}_3,\vec{k}_4)\;,
\end{equation}
with:
\begin{align}
\mathcal{M}_4(\vec{k}_1,\vec{k}_2,\vec{k}_3,\vec{k}_4) = & -2\frac{k_1^2k_3^2}{k_{12}^2k_{34}^2}\frac{W_{24}}{k_t}\left( \frac{\vec{Z}_{12}\cdot \vec{Z}_{34}}{k_{34}^2} +2\vec{k}_2\cdot \vec{Z}_{34} +\frac{3}{4}\sigma_{12}\sigma_{34} \right) - \nonumber \\
& - \frac{1}{2} \frac{k_3^2}{k_{34}^2}\sigma_{34}\left( \frac{\vec{k}_1\cdot \vec{k}_2}{k_t}W_{124} + 2\frac{k_1^2k_2^2}{k_t^3} + 6\frac{k_1^2k_2^2k_4}{k_t^4} \right)\;,
\end{align}
\begin{equation}
\sigma_{ab}=\vec{k}_a\cdot \vec{k}_b + k_b^2\;,
\end{equation}
\begin{equation}
\vec{Z}_{ab}=\sigma_{ab}\vec{k}_a - \sigma_{ba}\vec{k}_b\;,
\end{equation}
\begin{equation}
W_{ab}=1+\frac{k_a+k_b}{k_t}+\frac{2k_ak_b}{k_t^2}\;,
\end{equation}
\begin{equation}
W_{abc}=1+\frac{k_a+k_b+k_c}{k_t}+\frac{2(k_ak_b+k_bk_c+k_ak_c)}{k_t^2} + \frac{6k_ak_bk_c}{k_t^3}\;.
\end{equation}

It is easy to verify that the first term in brackets in eq.~\eqref{eq:Malda} has vanishing conformal variation and thus does not contribute to eq.~\eqref{eq:masterzeta}. This must be the case as the 4-point function does not depend on $\eta$. The right-hand side of the conformal consistency relation therefore reads
\be
\frac{H^6}{8\epsilon^3 M_P^6} \frac{1}{2 q^3\prod (2k_i^3)} \left(-\frac\epsilon2\right) \sum_{a=1}^{3} \vec{q}\cdot \vec{k}_a \left[ \frac{4}{k_a}\frac{\partial}{\partial k_a} + \frac{\partial^2}{\partial k_a^2} \right] \left(\sum_{i\neq j}k_i k_j^2 + \frac{8}{k_t}\sum_{i>j} k_i^2 k_j^2 \right) \;.
\ee
It is straightforward to verify that this indeed reproduces the $1/q^2$ behavior of the 4-point function \eqref{eq:4pfsr}.

\section{\label{sec:tilt}Including the deviation from scale invariance}
So far we have neglected the deviation from scale invariance, which gives the leading squeezed limit contribution $\sim 1/q^3$. The inclusion of the tilt is not entirely straightforward, as it is not obvious how it will affect the $1/q^2$ corrections.  Let us go through our arguments once again without assuming scale invariance. The starting point is that the effect of a background mode $\zb$ on an $n$-point function, at zeroth and first order in the gradients of $\zb$, just reduces to an appropriate transformation of the spatial coordinate: a dilation and a special conformal transformation. In mathematical terms
\be
\langle \zeta(\vec{x}_1) \ldots \zeta(\vec{x}_n) \rangle_{\zb\;,\nabla\zb} = \langle \zeta(\tilde{\vec{x}}_1) \ldots \zeta(\tilde{\vec{x}}_n) \rangle_0
\ee
with
\be
\label{eq:conftilde}
\tilde{\vec{x}}_i = \vec{x}_i + \zb(\vec{x}_+) \vec{x}_i + (\vec{x}_i - \vec{x}_+) (\nabla\zb \cdot (\vec{x}_i - \vec{x}_+)) - \frac12 (\vec{x}_i - \vec{x}_+)^2 \nabla\zb \;.
\ee
The transformation is done at the midpoint $\vec x_+ \equiv (\vec x_1 + \ldots + \vec x_n)/n$ although, as we will see later, this choice does not affect the result. The variation of the $n$-point function is at linear order in $\zb$:

\begin{align}
& \left(\overline{\zeta}(\vec {x}_+)\frac{\partial}{\partial\overline{\zeta}} + \nabla\zb \frac{\partial}{\partial \nabla\zb} \right)\int\frac{\rmd^3\vec{k}_1}{(2\pi)^3} \ldots \frac{\rmd^3\vec{k}_n}{(2\pi)^3} \langle \zeta_{\vec{k}_1} \ldots \zeta_{\vec{k}_n} \rangle_0 \;e^{ i \vec{k}_1 \tilde{\vec{x}}_1 +\cdots +i\vec{k}_n \tilde{\vec{x}}_n } = \nonumber \\  & \left(\overline{\zeta}(\vec {x}_+)\frac{\partial}{\partial\overline{\zeta}} + \nabla\zb \frac{\partial}{\partial \nabla\zb} \right)\int\frac{\rmd^3\vec{k}_1}{(2\pi)^3} \ldots \frac{\rmd^3\vec{k}_n}{(2\pi)^3} \widetilde{\langle \zeta_{\vec{k}_1} \ldots \zeta_{\vec{k}_n} \rangle_0} \;e^{ i \vec{k}_1 {\vec{x}}_1 +\cdots +i\vec{k}_n {\vec{x}}_n  } \;,
\end{align}
where the tilde on the correlation function indicates the action of the dilation and special conformal transformation in Fourier space. To get to the $(n+1)$-point function, we have to multiply by $\zb(\vec x)$ and average over the long mode. Let us consider the contribution from $\zb$ and $\nabla\zb$ separately. The first one gives
\begin{align}
\label{eq:dilation}
\langle  \zb(\vec{x}) \zeta(\vec{x}_1) \ldots \zeta(\vec{x}_n) \rangle \supset & \int \frac{\rmd^3\vec{q}}{(2\pi)^3} \frac{\rmd^3\vec{k}_1}{(2\pi)^3} \ldots \frac{\rmd^3\vec{k}_n}{(2\pi)^3} P(q) \left[- 3(n-1) - \sum_a \vec k_a \cdot \vec\partial_{k_a} \right] \langle \zeta_{\vec{k}_1} \ldots \zeta_{\vec{k}_n} \rangle'_0 \nonumber \\ & (2\pi)^3 \delta(\vec{k}_1+\cdots +\vec{k}_n ) e^{ i \vec{k}_1 {\vec{x}}_1 +\cdots +i\vec{k}_n {\vec{x}}_n  } e^{ i \vec{q} (\vec x -\vec{x}_+)} \;.
\end{align}
The terms proportional to $\nabla \zb$ on the right hand side of eq.~\eqref{eq:conftilde} can be expanded to  give transformations around the origin
\be
\label{eq:expand}
\vec{x}_i  (\nabla\zb \cdot \vec{x}_i ) - \frac12 \vec{x}_i^2 \nabla\zb - \vec{x}_i (\nabla\zb \cdot \vec{x}_+) - \vec{x}_+ (\nabla\zb \cdot \vec{x}_i ) + (\vec{x}_i \cdot \vec{x}_+) \nabla\zb + \vec{x}_+  (\nabla\zb \cdot \vec{x}_+ ) - \frac12 \vec{x}_+^2 \nabla\zb\;.
\ee
The last two terms describe a translation: this does not affect the $n$-point function which is translationally invariant. Analogously, the third-to-last term together with the fourth-to-last describe a rotation, which again does not change the correlation function. The third term describes an additional dilation and gives a contribution to the $(n+1)$-point function of the form
\begin{align}
\langle  \zb(\vec{x}) \zeta(\vec{x}_1) \ldots \zeta(\vec{x}_n) \rangle \supset & \int \frac{\rmd^3\vec{q}}{(2\pi)^3} \frac{\rmd^3\vec{k}_1}{(2\pi)^3} \ldots \frac{\rmd^3\vec{k}_n}{(2\pi)^3} P(q) \left[-3(n-1) - \sum_a \vec k_a \cdot \vec\partial_{k_a}\right] \langle \zeta_{\vec{k}_1} \ldots \zeta_{\vec{k}_n} \rangle'_0 \nonumber \\ & (2\pi)^3 \delta(\vec{k}_1+\cdots +\vec{k}_n ) e^{ i \vec{k}_1 {\vec{x}}_1 +\cdots +i\vec{k}_n {\vec{x}}_n  } e^{ i \vec{q} (\vec x - \vec x_+)} (i \vec q \cdot \vec x_+) \;.
\end{align}
This expression cancels, up to terms of order $q^2$, the dependence on $x_+$ in eq.~\eqref{eq:dilation}. This shows that the choice of the point $x_+$ is immaterial.

We now come to the action of the special conformal transformation, i.e.~the first two terms of eq.~\eqref{eq:expand}. To derive the action of these in Fourier space we have to convert powers of $x_i$ into derivatives with respect to momenta and integrate these by parts. In doing this, some derivatives act on the $\delta$-function of momentum conservation, as explained in \cite{arXiv:1104.2846}, and give a term with a distribution $\nabla \delta$ multiplied by the dilation transformation of the correlation function. This gives
\begin{align}
\langle  \zb(\vec{x}) \zeta(\vec{x}_1) \ldots \zeta(\vec{x}_n) \rangle \supset & \int \frac{\rmd^3\vec{q}}{(2\pi)^3} \frac{\rmd^3\vec{k}_1}{(2\pi)^3} \ldots \frac{\rmd^3\vec{k}_n}{(2\pi)^3} P(q) \left[-3(n-1) - \sum_a \vec k_a \cdot \vec\partial_{k_a}\right] \langle \zeta_{\vec{k}_1} \ldots \zeta_{\vec{k}_n} \rangle'_0 \nonumber \\ & \vec q \cdot (2\pi)^3 \nabla \delta(\vec{k}_1+\cdots +\vec{k}_n ) e^{ i \vec{k}_1 {\vec{x}}_1 +\cdots +i\vec{k}_n {\vec{x}}_n  } e^{ i \vec{q}  (\vec x - \vec x_+)}  \;.
\end{align}
Notice that the term $\vec q \cdot \vec x_+$ can be dropped as it gives contributions of order $q^2$.
If we sum the expression above with eq.~\eqref{eq:dilation} we end up with the full momentum conservation $(2\pi)^3 \delta(\vec q + \vec k_1 +\ldots +\vec k_n)$. Besides these singular terms we also have the standard special conformal transformation in Fourier space. In total we get to
\be
\boxed{
\begin{split}\label{eq:mastertilt}
\langle \zeta_{\vec q} \zeta_{\vec k_1} \ldots \zeta_{\vec k_n} \rangle'  \overset{q\rightarrow 0}{=}  &  - P(q) \left[3(n-1)  +\sum_a \vec k_a \cdot \vec\partial_{k_a} + \frac12  q^i D_i \right] \langle \zeta_{\vec{k}_1} \ldots \zeta_{\vec{k}_n} \rangle'  + {\cal{O}}(q/k)^2   \\ & \quad {\rm with } \quad q^iD_i \equiv \sum_{a=1}^{n} \Big[ 6\vec{q}\cdot \vec{\partial}_{k_a} - \vec{q}\cdot \vec{k}_a \vec{\partial}_{k_a}^2 +2\vec{k}_a \cdot \vec{\partial}_{k_a} (\vec{q}\cdot \vec{\partial}_{k_a}) \Big]  \;.
\end{split}}
 \ee
 Notice that the procedure above implies that the scale transformation of the $n$-point function is evaluated exactly with the momenta $\vec k_1 \ldots \vec k_n$ that appear on the left-hand side, even if at first order in $q$ these will not form a closed polygon. This is irrelevant for the conformal transformation which is already at order $q$. Equation \eqref{eq:mastertilt} represents the most general result of this paper.

\subsection{Resonant non-Gaussianities}
Models with a periodic modulation of the inflaton potential \cite{Chen:2008wn} have recently attracted attention, as motivated by explicit string constructions of Monodromy Inflation \cite{McAllister:2008hb}. For our purposes, these models represent an interesting example in which the $n$-point functions deviate sizably from scale-invariance---so that the results of this Section come into play---and in which many $n$-point functions, and not only the bi- and trispectra, can be analytically calculated \cite{Leblond:2010yq, arXiv:1111.3373}. 

The $(n+1)$-point function reads\footnote{This result holds in the decoupling limit and neglecting diagrams with more than one vertex. Corrections to the decoupling limit are suppressed by $\epsilon$ and they vanish in the limit $\epsilon \to 0$ keeping constant the power spectrum of $\zeta$ \cite{arXiv:1111.3373}. Terms with more than one vertex contain extra powers of the amplitude of the potential modulation. These corrections eventually become sizeable for $n$-point functions with large $n$ \cite{Leblond:2010yq}. However, the consistency relations must be satisfied independently by the single-vertex diagrams, as they are the only ones proportional to the first power of the amplitude of the modulation. The integral over $\eta$ can be explicitly done in the saddle-point approximation, in the limit of many oscillations per Hubble time \cite{Leblond:2010yq, arXiv:1111.3373}. Since the consistency relations must hold independently of this limit, and not only for oscillatory but for general potentials, here we prefer to keep the integral form.} \cite{Leblond:2010yq, arXiv:1111.3373}
\be
\langle \zeta_{\vec q} \zeta_{\vec k_1} \ldots \zeta_{\vec k_n} \rangle = (2\pi)^3\delta(\vec q +\vec k_1 + \ldots + \vec k_n) \left(-\frac{H}{\dot\phi}\right)^{n+1}\frac{H^{2n-2}}{2q^3\prod_{i=1}^n 2k_i^3} I_{n+1}\;,
\ee
where $I_{n+1}$ is given by the following in-in integral:
\be
I_{n+1}=- 2 \,\mathrm{Im} \int_{-\infty- i \epsilon}^0 \frac{\rmd \eta}{\eta^4} V^{(n+1)}(\phi(\eta)) (1-iq\eta)(1-ik_1\eta)\ldots(1-ik_n\eta)e^{ik_t\eta} \;.
\ee
Here $k_t = q + \sum_i k_i$  and $V^{(n+1)}$ is the $(n+1)$th derivative of the potential, evaluated on the unperturbed history $\phi(\eta)$. In resonant models, one starts from a trigonometric function and therefore these derivatives are simply sines or cosines. Here we keep the discussion more general, without specifying the form of the potential. 

Let us verify our general consistency relation eq.~\eqref{eq:mastertilt}.  First, notice that the action of the operator $q^i D_i$ on the $n$-point function vanishes. Given that the function only depends on the moduli of the wavevectors, the differential operator takes the simplified form \eqref{eq:confsimple}
\be
\sum_{a=1}^n \vec{q}\cdot \vec{k}_a \left[ \frac{4}{k_a}\frac{\partial}{\partial k_a} + \frac{\partial^2}{\partial k_a^2} \right] \langle \zeta_{\vec k_1} \ldots \zeta_{\vec k_n} \rangle' = 0\;.
\ee
To see that this expression vanishes, it is enough to compute the action of the operator in brackets for one particular momentum $k_a$. The result does not depend on $a$ (indeed the form of the amplitude remains the same with an additional $-\eta^2$ that appears in the integral). This means that the sum will be proportional to $\vec{q}\cdot \sum_{a=1}^n \vec{k}_a$, which is zero up to $q^2$ corrections.

This result seems to imply that correlation functions in these models are invariant under special conformal transformations. This cannot be true because correlation functions are not scale-invariant, due to the features in the potential, and there is no such thing as invariance under special conformal transformations without scale-invariance (indeed the commutator of a special conformal transformation with a translation contains a dilation). As we discussed above, there is an additional singular piece of the special conformal transformation, proportional to the derivative of the delta function and to the scale variation of the $n$-point function---which is not zero in this case---and this was used to derive eq.~\eqref{eq:mastertilt}. The vanishing of the conformal operator in Fourier space tells us that in these models, where there is no operator with spatial derivatives that enters at order $q$ in the consistency relation, the conformal transformation of the $n$-point function is just fixed by its dilation properties. 

Let us therefore check that the dilation transformation captures the squeezed limit of the $n$-point function up to corrections ${\cal O}(q^2)$.
In the squeezed limit $q \to 0$, the expression for the $(n+1)$-point function gives
\be
\langle \zeta_{\vec q} \zeta_{\vec k_1} \ldots \zeta_{\vec k_n} \rangle_{q\rightarrow 0}' = -  \left(-\frac{H}{\dot\phi}\right) \frac{H^2}{2q^3} \left(-\frac{H}{\dot\phi}\right)^{n}\frac{H^{2n-4}}{\prod_{i=1}^n 2k_i^3} 2\,\mathrm{Im} \int_{-\infty}^0 \frac{\rmd \eta}{\eta^4} V^{(n+1)} \prod_{i=1}^n(1-ik_i\eta)e^{ik_i\eta}\;,
\ee
up to ${\cal O}(q^2)$ corrections.
Now we can convert the $(n+1)$th derivative of the potential using
\be
\frac{d}{d\phi} V^{(n)}(\phi(\eta)) = - \frac{H}{\dot\phi} \left[\eta \frac{d}{d\eta} V^{(n)}(\phi(\eta)) \right] 
\ee
and integrate by parts the derivative with respect to $\eta$. This derivative can be traded for derivatives with respect to the $k_i$ to give
\begin{align}
& -  \left(\frac{H}{\dot\phi}\right)^2 \frac{H^2}{2q^3} \left[3(n-1) + \sum_i k_i \frac{\partial}{\partial k_i} \right] \left(-\frac{H}{\dot\phi}\right)^{n}\frac{(-1)H^{2n-4}}{\prod_{i=1}^n 2k_i^3} 2 \,\mathrm{Im} \int_{-\infty}^0 \frac{\rmd \eta}{\eta^4} V^{(n)} \prod_{i=1}^n(1-ik_i\eta)e^{ik_i\eta} \nonumber \\ & = - P(q) \left[3(n-1) + \sum_i k_i \frac{\partial}{\partial k_i} \right] \langle \zeta_{\vec{k}_1} \ldots \zeta_{\vec{k}_n} \rangle' \;,
\end{align}
as we wanted to show.

%

\section{\label{sec:tensors}Consistency relations for tensor modes}

In the previous Sections we discussed the generalization of the consistency relations involving scalar modes. There are similar relations involving tensor modes at zeroth order in the gradients \cite{Maldacena:2002vr}, and it is thus natural to wonder whether they can be generalized as well. We will be able to do so only in the case of one long-wavelength tensor mode, with the other legs being scalar modes. In all other cases, one is not able---at first order in gradients---to create the long mode with a coordinate transformation. For example, if one starts from the flat Euclidean metric, a special conformal transformation induces a conformal factor with a linear dependence on $x$, but the same will not work if one starts in the presence of a tensor mode.

Let us consider the $(n+1)$-point function with one background tensor mode $\langle \gamma^s_{\vec q}\zeta_{\vec k_1}\cdots \zeta_{\vec k_n} \rangle$ (\footnote{We are using the convention of \cite{Maldacena:2002vr} for the tensor modes.}). We work in the usual gauge, where the spatial line element is given by $a^2e^{2\zeta}\left( e^\gamma \right)_{ij}\mathrm{d}x_i\mathrm{d}x_j$ and $\gamma_{ij}$ is transverse and traceless \cite{Maldacena:2002vr,arXiv:1104.2846}. We want to induce a gravitational wave in the presence of scalar modes only, i.e.~starting with a line element $a^2e^{2\zeta}\delta_{ij}\mathrm{d}x_i\mathrm{d}x_j$. To induce a long-wavelength tensor mode, we consider a quadratic transformation of coordinates 
\be
x_i\rightarrow x_i + A_{ij}x_j + B_{ijk}x_jx_k \;,
\ee
where $B_{ijk}$ is symmetric in the last two indices $B_{ijk}=B_{ikj}$ (notice this has nothing to do with a conformal transformation). With this transformation the spatial line element becomes:
\be
a^2e^{2\zeta}\delta_{ij}\rmd x_i \rmd x_j \rightarrow a^2e^{2\zeta} \left( \delta_{ij} + 2A_{ij} + 4\frac{B_{ijk}+B_{jik}}{2}x_k \right) \rmd x_i \rmd x_j \;.
\ee
We can identify the terms in brackets with the gravitational wave; imposing the symmetry properties of $B_{ijk}$ we find:
\be
\label{eq:GWtrans}
A_{ij} = \frac{1}{2} \gamma_{ij} \;, \quad B_{ijk}=\frac{1}{4}(\partial_k \gamma_{ij} - \partial_i \gamma_{jk} + \partial_j \gamma_{ik}) \;.
\ee
So far we did not put any constraints on $\gamma_{ij}$. From the expressions for $A_{ij}$ and $B_{ijk}$ we see that we can consistently impose that $\gamma_{ij}$ is transverse and traceless. Now that we have the coordinate transformation that induces the long-wavelength tensor mode, we can follow the same logic as with scalar modes:
\be
\langle \zeta(\vec x_1)\ldots\zeta(\vec x_n) \rangle_\gamma = \langle \zeta(\tilde{\vec x}_1)\ldots\zeta(\tilde{\vec x}_n) \rangle \;,
\ee 
where transformation of coordinates is now given by:
\begin{align}
\label{eq:gravtransf}
\tilde{x}_i &= x_i + A_{ij}(\vec x_+) x_j + B_{ijk}(\vec x_+)(x_j-x^+_j)(x_k-x^+_k)  \nonumber \\
& = x_i + (A_{ij}(\vec x_+) - 2B_{ijk}(\vec x_+)x_k^+)x_j + B_{ijk}(\vec x_+)x_jx_k+ \mbox{translation} \;.
\end{align}

We end up, leaving the details of the calculation to Appendix \ref{app:GW}, with
\begin{align}
\label{eq:mastergrav}
\langle \gamma_{\vec q}^s & \zeta_{\vec k_1}\ldots\zeta_{\vec k_n} \rangle'_{q\rightarrow 0} =  - \frac{1}{2} P_\gamma(q) \sum_a \epsilon_{ij}^{s}(\vec{q}) k_{ai}  \partial_{k_{aj}} \langle  \zeta_{\vec k_1}\ldots \zeta_{\vec k_n} \rangle' \nonumber \\
& - \frac{1}{4} P_\gamma(q) \sum_a \epsilon_{ij}^{s}(\vec{q}) \left( 2 k_{ai} (\vec{q}\cdot \vec{\partial}_{k_a}) - (\vec{q}\cdot \vec{k_a}) \partial_{k_{ai}} \right) \partial_{k_{aj}} \langle  \zeta_{\vec k_1}\ldots \zeta_{\vec k_n} \rangle' \;.
\end{align}
This is the generalized consistency relation for the $(n+1)$-point function with $n$ scalar modes and a long background gravitational wave. The graviton power spectrum is given by $P_\gamma(q)=\frac{H^2}{M_P^2} \cdot \frac{1}{q^3}$. As before, the sum of the momenta entering in the $n$-point function is not exactly zero and this will produce the corrections linear in $\vec q$ coming from dilations.

We can check this equation in the simple case of the 3-point function \cite{Maldacena:2002vr}
\be
\langle \gamma_{\vec{q}}^s \zeta_{\vec{k}_1} \zeta_{\vec{k}_2} \rangle' =  \frac{H^4}{q^3k_1^3k_2^3}\frac{1}{4\epsilon}  \epsilon_{ij}^{s}(\vec q) k_{1i} k_{2j} \left( -k_t + \frac{k_2q+ k_1q+ k_2k_1}{k_t} + \frac{q k_1 k_2}{k_t^2} \right) \;.
\ee
In the limit $\vec q \rightarrow 0$ the amplitude becomes:
\be
\label{eq:gravlimit}
\langle \gamma_{\vec{q}}^s \zeta_{\vec{k}_1} \zeta_{\vec{k}_2} \rangle_{q\rightarrow 0} = \frac{H^2}{4\epsilon k_1^3}\frac{H^2}{q^3} \epsilon_{ij}(\vec q) \frac{k_{1i}k_{1j}}{k_1^2} \frac{3}{2} \left( 1 - \frac{5}{2}\frac{\vec k_1 \cdot \vec q}{k_1^2} \right) \;.
\ee
This coincides with the right-hand side of the consistency relation, eq.~\eqref{eq:mastergrav}. Notice that the second line of eq.~\eqref{eq:mastergrav} vanishes and that linear corrections come from evaluating the 2-point function with the sum of the two momenta not exactly equal to zero.

Let us come back to the question of whether it is possible to generalize the consistency relations to cases where we have some short gravitational waves. In this case we start from a line element of the form $a^2e^{2\zeta}\left( e^\gamma \right)_{ij}\mathrm{d}x_i\mathrm{d}x_j$. If one wants to consider the effect of a long-wavelength mode of $\zeta$ it is still possible to induce it, at zeroth order in gradients, by a dilation of the spatial coordinates. But a special conformal transformation induces a conformal factor only if one starts from a metric proportional to $\delta_{ij}$ and not in general. It is thus not possible to induce a constant gradient of $\zeta$ to generalize the consistency relations. The same arguments apply when considering a long tensor mode. The transformation eq.~\eqref{eq:GWtrans} which induces a long tensor mode including the first gradient corrections does not work in the presence of short tensor modes, as now also the part which is antisymmetric in $ij$, which drops out starting from $\delta_{ij}$, gives a contribution.

All we said can be seen at the level of the action as well. For example, looking at the cubic action for one tensor mode and two scalar modes in a slow-roll model
\be
S=\int d^4 x \;\frac{1}{2}\frac{\dot{\phi}^2}{H^2}a\gamma_{ij}\partial_i\zeta \partial_j \zeta\;,
\ee
we can see that the effect of $\gamma$ is equivalent, including first gradient corrections, to the spatial change of coordinates discussed above. A similar analysis does not work for the vertices $ \zeta_L \gamma \gamma$ and $\gamma_L \gamma \gamma$ when gradient corrections are included.

\section{\label{sec:internal}Soft internal lines}

So far we have considered squeezed limits, in which one of the external momenta goes to zero. However, using similar arguments, one can discuss the case when the sum of $m$ (out of $n$) of the external momenta goes to zero. In this case the $n$-point function is dominated by the exchange of a soft---small momentum---intermediate state. The result factorizes in the product of the power spectrum of the soft internal line and the $m-$ and $(n-m)$-point functions, both calculated in the presence of the long mode \cite{Seery:2008ax,Leblond:2010yq}. Indeed, the soft mode---scalar or tensor---freezes much before the others and behaves as a classical background. Following the arguments of this paper, it is straightforward to take into account a constant gradient of the long mode, i.e.~the first correction in its momentum. In mathematical terms
\begin{align}
\label{eq:cclimit}
\langle \zeta_{\vec k_1}\cdots \zeta_{\vec k_n} \rangle'_{\vec q\rightarrow 0} &= P_\zeta(q)\langle \zeta_{-\vec q}\zeta_{\vec k_1}\cdots \zeta_{\vec k_m} \rangle^*_{\vec q\rightarrow 0} \langle \zeta_{\vec q} \zeta_{\vec k_{m+1}}\cdots \zeta_{\vec k_n} \rangle^*_{\vec q\rightarrow 0} + \nonumber \\
 & \quad + P_\gamma(q)\sum_s \langle \gamma^s_{-\vec q}\zeta_{\vec k_1}\cdots \zeta_{\vec k_m} \rangle^*_{\vec q\rightarrow 0} \langle \gamma^s_{\vec q} \zeta_{\vec k_{m+1}}\cdots \zeta_{\vec k_n} \rangle^*_{\vec q\rightarrow 0}\;,
\end{align}
where $1<m<n$, $\vec q=\vec k_1 +\cdots +\vec k_m$ and the $*$ on the squeezed-limit correlation functions means that both the delta function of momentum conservation and the long-mode power spectrum $P(q)$ are dropped. Notice that, as a consequence of momentum conservation, the vector $\vec q$ has opposite sign in the two terms of the product. We can then evaluate the equation above using our previous results eq.~\eqref{eq:mastertilt} and eq.~\eqref{eq:mastergrav}.

The simplest example in which an internal line becomes soft is the 4-point function in the limit in which two of the external momenta become equal and opposite (counter-collinear limit) \cite{Seery:2008ax,Leblond:2010yq}. In the case of scalar exchange we know that the 3-point function $\langle \zeta_{\vec q}\zeta_{\vec k_1}\zeta_{\vec k_2} \rangle$ does not have any linear corrections. This implies that, in the limit of exact scale-invariance, we have a $q^2$ suppression at each of the two vertices (this is indeed verified by the explicit results \cite{Chen:2009bc,Arroja:2009pd}) that completely kills the $q \to 0$ divergence from the power spectrum, so that the result is not dominated by the soft scalar exchange. The counter-collinear limit is therefore regular
\be
\langle \zeta_{\vec{k}_1} \zeta_{\vec{k}_2} \zeta_{\vec{k}_3} \zeta_{\vec{k}_4} \rangle'_{\vec q\rightarrow 0} \sim {\rm const.}
\ee

More interesting is the case of graviton exchange, as in general we have linear corrections. For example, plugging in eq.~\eqref{eq:cclimit} the result eq.~\eqref{eq:gravlimit} of the previous Section for slow-roll inflation, we have the graviton exchange in the counter-collinear limit ($\vec q = \vec k_1 + \vec k_2$, $\vec k_1\approx -\vec k_2$ and $\vec k_3 \approx -\vec k_4$):  
\be
\langle \zeta_{\vec{k}_1} \zeta_{\vec{k}_2} \zeta_{\vec{k}_3} \zeta_{\vec{k}_4} \rangle'^{GE}_{\vec q\rightarrow 0} = \frac{H^2}{q^3} \frac{H^2}{4\epsilon k_1^3} \frac{H^2}{4\epsilon k_3^3} \sum_s \epsilon^{s*}_{ij}(\vec q) \epsilon^s_{lm}(\vec q) \frac{9}{4}\frac{k_{1i}k_{1j}}{k_1^2} \frac{k_{3l}k_{3m}}{k_3^2} \left( 1 + \frac{5}{2}\frac{\vec k_1 \cdot \vec q}{k_1^2} -\frac{5}{2}\frac{\vec k_3 \cdot \vec q}{k_3^2} \right) \;.
\ee
We can see that our results allow to capture also the ${\cal O}(q)$ corrections in the counter-collinear limit, and this is exactly the result that one finds taking the appropriate limit of the full 4-point function induced by the graviton exchange \cite{Seery:2008ax}. 

We conclude by noticing that as the squeezed limits are important for observables which involve a large separation of scales (the scale-dependent bias for example), the same occurs in the case of soft internal lines. For example, the recently discussed power spectrum of the CMB $\mu$ distortion \cite{Pajer:2012vz} is sensitive to the 4-point function in the counter-collinear limit, and therefore very suppressed in all single-field models.

\section{\label{sec:conclusions}Conclusions and outlook}
In this paper we extended the single-field consistency relations in the limit in which one of the momenta---external or internal---goes to zero, to include the first correction in the soft momentum. We proved that this correction is related to the variation of the correlator under a special conformal transformation.

These ``soft" limits are phenomenologically relevant, as some of the proposed experimental tests of non-Gaussianity involve a separation of scales, and therefore probe the primordial correlation functions in one of these limits, where our results apply. For example, we showed that in any single-field model the 4-point function generally goes as $1/q^2$ in the squeezed limit (assuming scale-invariance), but that the coefficient of this divergence is suppressed by $\sim 10^{-5}$ compared to the 3-point function and thus very small. Therefore the scale-dependent bias induced by the trispectrum (see \cite{Scoccimarro:2011pz} and refs. therein) will be suppressed in any single-field model.

From the theoretical point view our results, besides being a useful check of the calculations, fully explore the symmetry of primordial perturbations generated by single-field inflation. In a completely model-independent way,  {\em in any single-field model primordial correlation functions of $\zeta$ are endowed with an $SO(4,1)$ symmetry, with dilations and special conformal transformations non-linearly realized by $\zeta$}. Notice that this statement is not as trivial as the simple equivalence between a conformal transformation of the spatial coordinates and a change of the scale factor induced by $\zeta$. Indeed this equivalence must be valid during the whole period when perturbations are generated, in order to have the $SO(4,1)$ symmetry at the end of inflation. This holds, of course, only in single field models. It is worth noticing that no further extension of these consistency relations for scalar modes is possible: at second order in gradients, the long wavelength mode induces curvature and therefore cannot be removed by a clever change of coordinates. Similarly to what happens with the $1/q^3$ consistency relations, our eq.~\eqref{eq:mastertilt} encodes the fact that a long mode cannot be observed locally, until one is sensitive to its curvature.

It is worthwhile to stress that in this paper we switched between two different interpretations of the group $SO(4,1)$. In the decoupling limit it is the de Sitter isometry group while, later on, it is the conformal group of 3d Euclidean space. The first interpretation requires exact de Sitter, but it can be applied to test fields \cite{Creminelli:2011mw}, models with more than one field and to higher spins \cite{arXiv:1104.2846}. The second does hold in a general FRW, but it requires a single field model.

Let us conclude with some comments on possible future directions. Even if we phrased the paper using inflation as the key example, our results are more general and can be extended to other single-field models like \cite{ArmendarizPicon:2003ht,Khoury:2008wj,Khoury:2009my,Khoury:2011ii}.
It should also be possible to generalize some of the above results to multifield models, at least in the case of exact de Sitter geometry. The inflaton background spontaneously breaks this symmetry, so that the variation of a correlation function under the de Sitter isometry group should always be connected with the soft emission of an inflaton perturbation.  However, it is not clear whether this can be simply connected with observable quantities. Similarly one should be able to extend these relations to models where the $SO(4,1)$ symmetry is not the isometry group of de Sitter, but a subgroup of the 4d conformal group $SO(4,2)$, with additional constraints induced by the non-linear realization of $SO(4,2)$ \cite{Rubakov:2009np,arXiv:1007.0027,Hinterbichler:2011qk,Hinterbichler:2012mv}. 

Work is in progress by other groups to understand the consistency relations in terms of spontaneously broken symmetries \cite{JKL}, either in terms of Ward identities \cite{JKL2} or using path integrals and OPE arguments  \cite{NWL}.


\section*{Acknowledgments}
This paper was heavily influenced by discussions with W.~Goldberger, K.~Hinterbichler, L.~Hui, J.~Khoury and A.~Nicolis and it is thus related to the paper \cite{JKL} and additional work in progress \cite{JKL2, NWL}. We also thank G.~D'Amico, M.~Musso, E.~Pajer and L.~Senatore for useful discussions. JN is supported by FP7-IDEAS-Phys.LSS 240117.

\appendix
\section{\label{app:MD}Non-linear metric in matter dominance} 
In this Appendix we want to verify in an explicit example that the second-order metric in $\zeta$-gauge, with one of the modes much longer than the Hubble radius can be obtained simply as a coordinate transformation from the first-order metric, and that the result is accurate up to corrections of order $(k_L/a H)^2$, where $k_L$ is the long-wavelength momentum. We are going to study the second-order metric in matter dominance, written in terms of the initial perturbation $\zeta_0$, assuming no initial tensor mode. This has been studied in $\zeta$-gauge in \cite{Boubekeur:2008kn}.

The form of the line element is given by:
\be
ds^2 = -N^2 dt^2 +h_{ij}(d x^i + N^i dt)(d x^j + N^j dt) \;,\qquad h_{ij} = a^2(t) e^{2 \zeta} (\delta_{ij} + \gamma_{ij}) \;,
\ee
with $\gamma$ transverse and traceless. Setting $N_i = \partial_i\psi + N_{i}^T$ we have
\begin{eqnarray}
N & = & 1 \label{eq:2ndN}\\
\psi & = & -\frac{2}{5H}\z_0 + \frac{1}{5H}\partial^{-2}\[(\partial_j\z_0)^2-3\partial^{-2}\partial_j\partial_k(\partial_j\z_0\partial_k\z_0)\] \nonumber \\
& & + \frac{4}{25a^2H^3}\partial^{-2}\[\frac{3}{7}(\partial^2\z_0)^2 + \partial_i\z_0\partial_i\partial^2\z_0+\frac{4}{7}(\partial_i\partial_j\z_0)^2\]\;, \label{omega}\\
N_i^T & =& -\frac{4}{5 H} \partial^{-2}\left[\partial_i \zeta_0 \partial^2\zeta_0 -
\partial^{-2}\partial_i\partial_k (\partial_k\zeta_0 \partial^2\zeta_0)\right] \;,\label{trans}\\
\zeta & = & \z_0 -\frac{1}{5a^2H^2}\partial^{-2}\partial_i\partial_j(\partial_i\z_0\partial_j\z_0) \label{Psi} \;.
\label{gammaij}
\end{eqnarray}
Tensor modes $\gamma$ will also be induced by the initial scalar perturbations; their explicit form can be found in \cite{Boubekeur:2008kn}, but we will not need it here. We are interested in this metric in the limit in which one of the modes in the initial conditions, let us call it $\zeta_L$, is (for all times) much longer than the Hubble radius, keeping only terms which are at most linear in the gradients of this long mode. In particular we are interested in the quadratic terms with a long and a short mode, which we call $\zeta_S$. In the limit we are considering these second order terms reduce to
\begin{eqnarray}
\label{eq:2ndsqueezedNi}
N_i = \partial_i\psi + N_{i}^T & = & \frac{4}{25a^2H^3}\partial_j\zeta_L\partial_i\partial_j\zeta_S - \frac{4}{5 H}\zeta_S\partial_i\zeta_L \;, \\ \label{eq:2ndsqueezedzeta}
\zeta & = & -\frac{2}{5a^2H^2}\partial_i\zeta_L\partial_i\zeta_S \;.
\end{eqnarray}

The linear metric
\be
ds^2 = -dt^2 - \frac{4}{5 H} \partial_i\zeta_L dt dx^i + a^2(1+ 2 \zeta_L) d \vec x^2
\ee
can be obtained starting from the unperturbed FRW with a coordinate transformation. One has to perform a conformal transformation of the spatial coordinates, which multiplies the spatial metric by a constant term and a term linear in the coordinates: in this way one reproduces the spatial dependence of $\zeta_L$ up to corrections involving the second derivatives of $\zeta_L$.  To give the correct space-time component one has to make a time-dependent translation
\be
\label{eq:tdtrans}
\tilde x^i = x^i + \frac25  \frac{1}{a^2 H^2} \partial_i \zeta_L \;.
\ee
Now we want to apply the same coordinate transformation on the metric at first order perturbed by the short mode 
\be
\label{eq:1stshort}
ds^2 = -dt^2 - \frac{4}{5 H} \partial_i\zeta_S dt dx^i + a^2(1+ 2 \zeta_S) d \vec x^2
\ee
and reproduce eq.s~\eqref{eq:2ndsqueezedNi} and \eqref{eq:2ndsqueezedzeta}. Notice first of all that this change of coordinates does not change $N$, which indeed is unperturbed even before taking one of the modes to be super-Hubble, eq.~\eqref{eq:2ndN}. The first term on the right-hand side of \eqref{eq:2ndsqueezedNi} comes from the time-dependent translation \eqref{eq:tdtrans} of $\partial_i \zeta_S$ in eq.~\eqref{eq:1stshort} and the same happens for eq.~\eqref{eq:2ndsqueezedzeta}. The second term in \eqref{eq:2ndsqueezedNi} comes, on the other hand, from the spatial metric, when $dx$ is converted into $dt$ by the time-dependent translation. One can also check, from the explicit expression in  \cite{Boubekeur:2008kn}, that tensors are not produced in the limit in which we only keep the first gradients of the long mode.  

Notice that we have an additional set of second order terms coming from the fact that the short mode $\zeta_S$ is evaluated at the conformally transformed point and this transformation depends on $\zeta_L$. This is not a genuine non-linearity but a choice of initial conditions.

\section{\label{app:consaction}Consistency relations at the level of the action}
In this Appendix we  explicitly check that the cubic action for $\zeta$, in the limit in which one of the three modes is much longer than the other two, can be obtained with a coordinate transformation of the quadratic action, up to corrections which are {\em quadratic} in the long-wavelength momentum. This property of the action implies the conformal consistency relations for $\zeta$. We will focus on minimal slow-roll inflation, whose quadratic and cubic actions are given by  \cite{Maldacena:2002vr}
\be
S^{(2)} = \int \mathrm{d}^3 \vec x \mathrm{d}t\; \epsilon a^3 \left[ \dot{\zeta}^2 - \left(\frac{\partial\zeta}{a}\right)^2 \right]\;,
\ee
\begin{align}
S^{(3)}= & \int \mathrm{d}^3 \vec x \mathrm{d}t \left[ ae^\zeta \Big( 1+\frac{\dot{\zeta}}{H} \Big) \left( -2\partial^2\zeta-(\partial\zeta)^2 \right) + \epsilon a^3e^{3\zeta}  \dot{\zeta}^2 \Big( 1-\frac{\dot{\zeta}}{H} \Big) +\right. \nonumber \\
& \left. + a^3 e^{3\zeta} \left( \frac{1}{2}\left( \partial_i\partial_j\psi\partial_i\partial_j\psi - (\partial^2\psi)^2 \right) \Big( 1-\frac{\dot{\zeta}}{H} \Big) -2\partial_i\psi\partial_i\zeta\partial^2\psi  \right) \right]\;,
\end{align}
where $\psi= -\frac{\zeta}{a^2H} + \chi$, $\partial^2\chi = \epsilon \dot \zeta$ and we set $M_P =1$.
Let us consider the coordinate transformation that gives rise to the correct linear-order solution of the metric for the long-wavelength mode $\zeta_L$ 
\be
\label{eq:transfapp}
x^i \rightarrow \tilde{x}^i=x^i-b^ix^2+2x^i(b\cdot x) + \delta x^i(t)
\ee
with
\be
b_i = -\frac12 \partial_i\zeta_L \qquad \frac{d}{d t} \delta{x}^i(t) = -\frac{2b^i}{a^2H} + \mathcal{O}(\epsilon) \;.
\ee
The slow-roll correction in $\delta x^i$ is necessary to reproduce the $\chi$ contribution in $\psi$, but we will not need it below. Notice that we are here considering a change of coordinates which {\em induces} a long wavelength mode, while in the main text we consider a change of coordinates which {\em removes} the background $\zeta_L$: this is the source of a relative minus sign. The change of coordinates induces the following variation of the quadratic action
\be
\label{eq:varS2}
\delta S^{(2)}= \int \mathrm{d}^3 \vec x \mathrm{d}t \left[ -6 a^3\epsilon(\vec b\cdot \vec x)\dot{\zeta}^2 + 2 a\epsilon(\vec b\cdot \vec x)(\partial\zeta)^2 - 4a\epsilon \frac{\dot{\zeta}}{H}b_i \partial_i\zeta \right]\;.
\ee
Let us compare this with the cubic action at first order in gradients for the long mode.  We separate long and short modes by writing $\zeta=\zeta_L + \zeta_S$ and note that $\dot{\zeta}_L={\cal{O}}(k_L^2)$. Without loss of generality we are expanding $\zeta_L$ around the origin (the calculation for constant $\zeta_L$ and dilations can be found in \cite{Creminelli:2011rh}). The relevant terms from the expansion of the cubic action are those linear in $\zeta_L$. For example, the second term in $S^{(3)}$ gives
\be
\epsilon a^3e^{3\zeta} \dot{\zeta}^2 \Big( 1-\frac{\dot{\zeta}}{H} \Big) \rightarrow 3 \epsilon a^3 \zeta_L \dot{\zeta}_S^2 \;.
\ee
For the other contributions we have to do some integrations by parts. The first term in $S^{(3)}$ gives
\begin{align}
ae^\zeta \Big( 1+\frac{\dot{\zeta}}{H} \Big) \left( -2\partial^2\zeta-(\partial\zeta)^2 \right) & \rightarrow ae^{\zeta_L+\zeta_S} \Big( 1+\frac{\dot{\zeta}_S}{H} \Big) \left( -2\partial^2\zeta_S-(\partial\zeta_S)^2 -2\partial_i\zeta_L\partial_i\zeta_S \right) \nonumber \\
& \rightarrow -a\Big( - \zeta_L(\partial\zeta_S)^2 + 2\zeta_L\frac{\dot{\zeta}_S}{H} \partial^2 \zeta_S + 2\partial_i\zeta_L \frac{\dot{\zeta}_S}{H} \partial_i \zeta_S \Big) \;,
\end{align}
where we have done one integration by parts: $a\zeta_L\zeta _S\partial^2\zeta_S=-a\zeta_L(\partial\zeta_S)-a\partial_i\zeta_L \zeta_S \partial_i\zeta_S^2$. Similarly, we can transform the second term in brackets in the following way
\begin{align}
-2a\zeta_L\frac{\dot{\zeta}_S}{H} \partial^2 \zeta_S & \rightarrow  2a\partial_i\zeta_L \frac{\dot{\zeta}_S}{H}\partial_i \zeta_S  + \frac{a}{H} \zeta_L \frac{\mathrm{d}}{\mathrm{d}t}[(\partial\zeta_S)^2] \nonumber \\
& \rightarrow -a\Big( -2\partial_i\zeta_L \frac{\dot{\zeta}_S}{H} \partial_i \zeta_S + \zeta_L(\partial\zeta_S)^2 + \epsilon\zeta_L(\partial\zeta_S)^2  \Big) \;,
\end{align}
so that the final contribution from the first term in the cubic action is
\be
ae^\zeta \Big( 1+\frac{\dot{\zeta}}{H} \Big) \left( -2\partial^2\zeta-(\partial\zeta)^2 \right)  \rightarrow -a\epsilon\zeta_L(\partial\zeta_S)^2.
\ee
Finally, we can focus on the last term in $S^{(3)}$. It is easy to see that the relevant contribution will be given by
\be
\label{eq:S3psi}
\frac{3}{2}a^3 \zeta_L  \left( \partial_i\partial_j\psi_S\partial_i\partial_j\psi_S - (\partial^2\psi_S)^2 \right) - 2a^3\partial_i\psi_S\partial_i\zeta_L\partial^2\psi_S - 2a^3\partial_i\psi_L\partial_i\zeta_S\partial^2\psi_S \;.
\ee
We can see that the first term has the same form as the second one, once we integrate by parts. Moreover, both of these terms are total derivatives. This can be seen by applying the following identity
\be
\partial_i(\partial_if\partial_jf) = \partial^2f\partial_jf + \partial_if \partial_i\partial_j f = \partial^2f\partial_jf + \frac{1}{2}\partial_j(\partial f)^2 \;.
\ee
We are left only with the last contribution in eq.~\eqref{eq:S3psi}. We are forced to replace $\partial^2 \psi_S$ with $\partial^2\chi_S = \epsilon \dot\zeta_S$, otherwise we again get a total derivative term. At leading order in slow-roll we have $\partial_i\psi_L = -\frac{1}{a^2H}\partial_i\zeta_L$. Collecting all the relevant contributions we get
\be
\delta S^{(3)} = \int \mathrm{d}^3 \vec x \mathrm{d}t \left( 3 \epsilon a^3 \zeta_L \dot{\zeta}_S^2 -a\epsilon\zeta_L(\partial\zeta_S)^2 +\frac{2a\epsilon}{H} \dot{\zeta}_S\partial_i\zeta_S\partial_i\zeta_L \right) \;.
\ee 
This coincides with eq.~\eqref{eq:varS2} using $b_i = -\frac12 \partial_i\zeta_L$, as we wanted to show.

\section{\label{app:GW}Derivation of the consistency relation for tensor modes} 

In this Appendix we are going to give some details on the derivation of the consistency relation for short scalars and a long gravitational wave. We want to compute the average over the long gravitational wave $\gamma_{mn}(x)$ of the variation of the scalar $n$-point function induced by the change of coordinates given in eq.~\eqref{eq:gravtransf}. In order to simplify the notation let us define the following:
\be
C_{ij}(\vec x_+)=A_{ij}(\vec x_+) - 2B_{ijk}(\vec x_+)x_k^+, \quad \vec P = \sum_i\vec k_i, \quad \langle \zeta_{\vec k_1} \ldots \zeta_{\vec k_n} \rangle = (2\pi)^3\delta(\vec P) \mathcal{M}\;.
\ee
Under this change of coordinates, the transformation of the $n$-point function is given by
\be
\delta \langle \zeta(\vec x_1) \ldots \zeta(\vec x_n) \rangle = \sum_{a=1}^n \left(C_{ij}(\vec x_+)x_{aj}\partial_{ai} + B_{ijk}(\vec x_+)x_{aj} x_{ak} \partial_{ai} \right) \langle \zeta(\vec x_1) \ldots \zeta(\vec x_n) \rangle\;.
\ee
We can compute it term by term. The contribution from dilations is given by
\begin{align}
\delta\langle &\zeta(\vec x_1)\ldots\zeta(\vec x_n) \rangle \supset \sum_{a=1}^n C_{ij}(\vec x_+)x_{aj}\partial_{ai} \langle\zeta(\vec x_1)\ldots\zeta(\vec x_n) \rangle \nonumber \\
& =  \int \frac{\mathrm{d}^3 \vec k_1}{(2\pi)^3}\ldots \frac{\mathrm{d}^3 \vec k_n}{(2\pi)^3}(2\pi)^3 \sum_a C_{ij}(\vec x_+) \delta(\vec P)\mathcal{M} k_{ai} \partial_{k_{aj}} e^{i\vec k_1 \cdot \vec x_1 + \cdots + i\vec k_n \cdot \vec x_n} \nonumber \\
& =  \int \frac{\mathrm{d}^3 \vec k_1}{(2\pi)^3}\ldots \frac{\mathrm{d}^3 \vec k_n}{(2\pi)^3} (2\pi)^3\delta(\vec P) \left(-\sum_a C_{ij}(\vec x_+) k_{ai} \partial_{k_{aj}}\mathcal{M} \right)  e^{i\vec k_1 \cdot \vec x_1 + \cdots + i\vec k_n \cdot \vec x_n}\;.
\end{align}
Here we used the fact that the tensor $C_{ij}$ is traceless and that a derivative on the delta function can be rewritten as $\partial_{P_j}\delta(\vec P)$ leading to a term proportional to $P_i \delta(\vec P)$. Let us now use the explicit expression for $C_{ij}$:
\be
\label{eq:cij}
C_{ij}(\vec x_+ )= \frac{1}{2}(1-x_k^+\partial_k^+)\gamma_{ij}(\vec x_+) +  \frac{1}{2}\left(x_k^+\partial_i^+ \gamma_{jk}(\vec x_+) - x_k^+\partial_j^+ \gamma_{ik}(\vec x_+) \right)
\ee
and focus on the first term. Averaging over the long background mode $\gamma_{mn}(x)$ we get
\begin{align}
\langle \gamma_{mn}(x) & \zeta(\vec x_1)\ldots\zeta(\vec x_n) \rangle \supset \int \frac{\mathrm{d}^3 \vec q}{(2\pi)^3} \frac{\mathrm{d}^3 \vec k_1}{(2\pi)^3}\ldots \frac{\mathrm{d}^3 \vec k_n}{(2\pi)^3} (2\pi)^3\delta(\vec P) \left(-\sum_a k_{ai} \partial_{k_{aj}}\mathcal{M} \right) \nonumber \\
&  \times \sum_s\epsilon_{mn}^{s}(\vec q)\epsilon_{ij}^{s}(\vec q) \frac{H^2}{q^3} \frac{1}{2}(1-x_k^+\partial_k^+) e^{i\vec k_1 \cdot \vec x_1 + \cdots + i\vec k_n \cdot \vec x_n} e^{i\vec q \cdot \vec x - i\vec q \cdot \vec x_+}\;.
\end{align}
Notice that $(1-x_k^+\partial_k^+)e^{- i\vec q \cdot \vec x_+}=1+\mathcal{O}(q^2)$ which proves that there is no dependence of the result on a choice of the point $\vec x_+$ at first order in the small momentum $\vec q$. The second term of eq.~\eqref{eq:cij} has an additional dependence on $\vec x_+$ but it gives no contribution. After the averaging over the long background mode $\gamma_{mn}(x)$ the result is
\begin{align}
\langle \gamma_{mn}(x) & \zeta(\vec x_1)\ldots\zeta(\vec x_n) \rangle \supset \int \frac{\mathrm{d}^3 \vec q}{(2\pi)^3} \frac{\mathrm{d}^3 \vec k_1}{(2\pi)^3}\ldots \frac{\mathrm{d}^3 \vec k_n}{(2\pi)^3} (2\pi)^3\delta(\vec P) \left(-\sum_a k_{ai} \partial_{k_{aj}}\mathcal{M} \right) \nonumber \\
& \times \sum_s\epsilon_{mn}^{s}(\vec q) \frac{H^2}{q^3} \frac{1}{2}\left(iq_j\epsilon_{ik}^s(\vec q) - iq_i\epsilon_{jk}^s(\vec q) \right) x_k^+ e^{i\vec k_1 \cdot \vec x_1 + \cdots + i\vec k_n \cdot \vec x_n} e^{i\vec q \cdot \vec x - i\vec q \cdot \vec x_+}\;.
\end{align}
We can rewrite $x_k^+$ in front of the exponent like $i\partial_{q_k}$ and do a partial integration. In the first line in the integral nothing depends on $\vec q$. The action on the second line will produce three terms respectively proportional to
\begin{align}
& \sim \sum_s \left(\partial_{q_k}\epsilon_{mn}^{s}(\vec q)\right) \frac{1}{q^3} \left(q_j\epsilon_{ik}^s(\vec q) - q_i\epsilon_{jk}^s(\vec q) \right)\;,  \nonumber \\
& \sim \sum_s \epsilon_{mn}^{s}(\vec q) \frac{q_k}{q^5} \left(q_j\epsilon_{ik}^s(\vec q) - q_i\epsilon_{jk}^s(\vec q) \right)\;,  \\
& \sim \sum_s \epsilon_{mn}^{s}(\vec q) \frac{1}{q^3} \left(\epsilon_{ij}^s(\vec q) +q_j\partial_{q_k}\epsilon_{ik}^s(\vec q) - \epsilon_{ji}^s(\vec q) - q_i\partial_{q_k}\epsilon_{jk}^s(\vec q) \right) \;. \nonumber
\end{align}
The first term will vanish when we multiply both sides of the final equation by $\epsilon_{mn}^{*s}(\vec q)$ in order to express everything in terms of the amplitudes. It is easy to see that $\epsilon_{mn}^{*s}(\vec q)\partial_{q_k}\epsilon_{mn}^{s}(\vec q)=0$. The second term is trivially zero because polarization tensors are transverse. In order to prove that the last one is zero as well, let us first note that one can write the polarization tensor in the following form
\be
\epsilon_{ij}^{s_a}(\vec k_a)=\frac{1}{\sqrt{2}} (\epsilon^{P}_{ij}(\vec k_a)+\mathrm{i}s_a \epsilon^{X}_{ij}(\vec k_a))\;,
\ee
where 
\be
\epsilon^{P}_{ij}(\vec k_a) = z_iz_j - u^a_iu^a_j,\quad \epsilon^{X}_{ij}(\vec k_a) = z_i u^a_j + z_j u^a_i\;,
\ee
and unit vectors $\vec z$ and $\vec u_a$ are orthogonal to $\vec k_a$ and each other \cite{arXiv:1104.2846}. Using these expressions one can derive the following formula for the derivative of the polarization tensor
\be
\partial_{k_j}\epsilon^s_{lr}(\vec k)=-\frac{1}{k^2}\left( k_l\epsilon^s_{jr}(\vec k) + k_r\epsilon^s_{lj}(\vec k) \right)\;,
\ee
which one can use to complete the proof.

We can focus now on the second part of the coordinate transformation. The variation of the $n$-point function is
\begin{align}
\delta\langle &\zeta(\vec x_1)\ldots\zeta(\vec x_n) \rangle \supset \sum_{a=1}^n B_{ijk}(\vec x_+)x_{aj}x_{ak}\partial_{ai} \langle\zeta(\vec x_1)\ldots\zeta(\vec x_n) \rangle \nonumber \\
& =  \int \frac{\mathrm{d}^3 \vec k_1}{(2\pi)^3}\ldots \frac{\mathrm{d}^3 \vec k_n}{(2\pi)^3}(2\pi)^3 \sum_a (-i)B_{ijk}(\vec x_+) \delta(\vec P)\mathcal{M} k_{ai} \partial_{k_{aj}} \partial_{k_{ak}} e^{i\vec k_1 \cdot \vec x_1 + \cdots + i\vec k_n \cdot \vec x_n}  \nonumber \\
& = - \int \prod_b \frac{\mathrm{d}^3 \vec k_b}{(2\pi)^3} (2\pi)^3\sum_a iB_{ijk}(\vec x_+)k_{ai} \left( 2\partial_{k_{ak}}\delta({\vec P}) + \delta(\vec P) \partial_{k_{aj}} \right) \partial_{k_{ak}}\mathcal{M} e^{i\vec k_1 \cdot \vec x_1 + \cdots + i\vec k_n \cdot \vec x_n}\;,
\end{align}
where we have used similar tricks as before to get the final result. It is convenient to express $B_{ijk}$ as a sum of two terms:
\be
B_{ijk}(\vec x_+)= \frac{1}{4}\left(\partial_k^+\gamma_{ij}(\vec x_+)\right) - \frac{1}{4}\left( \partial_i^+ \gamma_{jk}(\vec x_+) - \partial_j^+ \gamma_{ik}(\vec x_+) \right)\;.
\ee
The term containing a derivative of the delta function in the variation of the $n$-point function together with the first term in $B_{ijk}$ give:
\begin{align}
\langle \gamma_{mn}(x) & \zeta(\vec x_1)\ldots\zeta(\vec x_n) \rangle \supset \int \frac{\mathrm{d}^3 \vec q}{(2\pi)^3} \frac{\mathrm{d}^3 \vec k_1}{(2\pi)^3}\ldots \frac{\mathrm{d}^3 \vec k_n}{(2\pi)^3} (2\pi)^3 \left(-\sum_a k_{ai} \partial_{k_{aj}}\mathcal{M} \right) \nonumber \\
& \times \vec q \cdot \nabla\delta(\vec P) \sum_s\epsilon_{mn}^{s}(\vec q)\epsilon_{ij}^{s}(\vec q) \frac{H^2}{q^3}\frac{1}{2}  e^{i\vec k_1 \cdot \vec x_1 + \cdots + i\vec k_n \cdot \vec x_n} e^{i\vec q \cdot \vec x - i\vec q \cdot \vec x_+}\;.
\end{align}
The form of this contribution is the same as for dilations. When they sum up the final result will be the same as for dilations but with a modified delta function $\delta(\sum \vec k_i + \vec q)$. This means that in all computations the $n$-point function has to be computed slightly off-shell. The two additional terms that contain a derivative of the delta function will give a total contribution proportional to
\be
\sim \partial_{P_k}\delta(\vec P) \epsilon_{jk}^s(\vec q) q_i \sum_a \left( k_{ai}\partial_{k_{aj}} - k_{aj}\partial_{k_{ai}} \right) \mathcal{M}\;.
\ee
This expression is equal to zero. In order to prove that, we can rewrite it taking into account that the amplitude is in general a function of scalar products $\vec k_a \cdot \vec k_b$
\be 
\sum_a \left( k_{ai}\partial_{k_{aj}} - k_{aj}\partial_{k_{ai}} \right) \mathcal{M} = \sum_{a,b=1}^n \frac{\partial \mathcal{M}}{\partial (\vec k_a \cdot \vec k_b)} \left( k_{bj} k_{ai} - k_{aj} k_{bi} \right) = 0\;.
\ee
Finally, the last part of the variation is
\be
\delta\langle \zeta(\vec x_1)\ldots\zeta(\vec x_n) \rangle \supset - \int \frac{\mathrm{d}^3 \vec k_1}{(2\pi)^3}\ldots \frac{\mathrm{d}^3 \vec k_n}{(2\pi)^3} (2\pi)^3\delta(\vec P)  \sum_a iB_{ijk}(\vec x_+)k_{ai}  \partial_{k_{aj}} \partial_{k_{ak}}\mathcal{M}  e^{i\vec k_1 \cdot \vec x_1 + \cdots + i\vec k_n \cdot \vec x_n}\;.
\ee
After averaging over the long mode it gives the following contribution
\begin{align}
\langle \gamma_{mn}(x) & \zeta(\vec x_1)\ldots\zeta(\vec x_n) \rangle \supset \int \frac{\mathrm{d}^3 \vec q}{(2\pi)^3} \frac{\mathrm{d}^3 \vec k_1}{(2\pi)^3}\ldots \frac{\mathrm{d}^3 \vec k_n}{(2\pi)^3} (2\pi)^3 \delta(\vec P) \left(-\sum_a k_{ai} \partial_{k_{aj}}\partial_{k_{ak}}\mathcal{M} \right) \nonumber \\
&  \sum_s\epsilon_{mn}^{s}(\vec q) \frac{H^2}{q^3}\frac{1}{4} \left(q_k\epsilon_{ij}^{s}(\vec q) - q_i\epsilon_{jk}^{s}(\vec q) +q_j\epsilon_{ik}^{s}(\vec q)\right)  e^{i\vec q \cdot \vec x + i\vec k_1 \cdot \vec x_1 + \cdots + i\vec k_n \cdot \vec x_n}\;.
\end{align}

Summing up all the nonzero terms and going to momentum space on the lefthand side of the equation, we get the generalized consistency relation for a background tensor mode:
\begin{align}
\langle \gamma_{\vec q}^s & \zeta_{\vec k_1}\ldots\zeta_{\vec k_n} \rangle'_{q\rightarrow 0} =  - \frac{1}{2} P_\gamma(q) \sum_a \epsilon_{ij}^{s}(\vec{q}) \left( k_{ai} + k_{ai} (\vec{q}\cdot \vec{\partial}_{k_a}) - \frac{1}{2}(\vec{q}\cdot \vec{k_a}) \partial_{k_{ai}} \right)  \partial_{k_{aj}} \langle  \zeta_{\vec k_1}\ldots \zeta_{\vec k_n} \rangle' \;.
\end{align}

\end{document}